\documentclass{aa}  

\usepackage{graphicx}
\usepackage{hyperref}

\usepackage{txfonts}
\usepackage{natbib}
\usepackage{multirow}

\newcommand{\lsi}{LS\,I\,+61$^\circ$\,303}
\newcommand{\unit}[1]{\mbox{\boldmath $\hat{#1}$}}
\newcommand{\rchi}{\chi}

\begin{document}

\title{Orbital variability of the optical linear polarization of the $\gamma$-ray binary \lsi\ and new constraints on the orbital parameters}

\titlerunning{Optical polarization in $\gamma$-ray binary \lsi}

\author{Vadim Kravtsov\inst{1,2}    
\and 
Andrei V. Berdyugin\inst{1}   
\and  
Vilppu Piirola\inst{1}   
\and 
Ilia A. Kosenkov\inst{1,3} 
\and
Sergey S. Tsygankov\inst{1,2}  
\and
Maria Chernyakova\inst{4,5}
\and
Denys Malyshev\inst{6}
\and 
Takeshi Sakanoi\inst{7}   
\and 
Masato Kagitani\inst{7}   
\and 
Svetlana V. Berdyugina\inst{8,9}   
\and
Juri Poutanen\inst{1,2,10} }

\authorrunning{V. Kravtsov et al.} 
 
\institute{Department of Physics and Astronomy,  FI-20014 University of Turku, Finland\\
\email{vakrau@utu.fi}
\and
Space Research Institute of the Russian Academy of Sciences, Profsoyuznaya Str. 84/32, Moscow 117997, Russia
\and
Department of Astrophysics, St. Petersburg State University, Universitetskiy pr. 28, Peterhof, 198504 St. Petersburg, Russia
\and
School of Physical Sciences and CfAR, Dublin City University, Dublin 9, Ireland
\and
Dublin Institute for Advanced Studies, 31 Fitzwilliam Place, Dublin 2, Ireland
\and
Institut f{\"u}r Astronomie und Astrophysik T{\"u}bingen, Universit{\"a}t T{\"u}bingen, Sand 1, D-72076 T{\"u}bingen, Germany
\and 
Graduate School of Sciences, Tohoku University, Aoba-ku,  980-8578 Sendai, Japan
\and
Leibniz-Institut f\"{u}r Sonnenphysik, Sch\"{o}neckstr. 6, 79104 Freiburg, Germany
\and
Institute for Astronomy, University of Hawaii, 2680 Woodlawn Drive, Honolulu, 96822-1897 HI, USA 
\and
Nordita, KTH Royal Institute of Technology and Stockholm University, Roslagstullsbacken 23, SE-10691 Stockholm, Sweden }


\abstract{We studied the variability of the linear polarization and brightness of the $\gamma$-ray binary \lsi.
High-precision \textit{BVR} photopolarimetric observations were carried out with the Dipol-2 polarimeter on the 2.2 m remotely controlled UH88 telescope at Mauna Kea Observatory and the 60 cm Tohoku telescope at Haleakala Observatory (Hawaii) over 140 nights in 2016--2019. 
We also determined the degree and angle of the interstellar polarization toward \lsi\ using two out of four nearby field stars that have \textit{Gaia}'s parallaxes. 
After subtracting the interstellar polarization, we determined the position angle of the intrinsic polarization $\theta \simeq 11{\degr}$, which can either be associated with the projection of the Be star's decretion disk axis on the plane of sky, or can differ from it by 90\degr. 
Using the Lomb-Scargle method, we performed timing analyses and period searches of our polarimetric and photometric data. 
We found statistically significant periodic variability of the normalized Stokes parameters $q$ and $u$ in all passbands. 
The most significant period of variability, $P_\text{Pol} = 13.244 \pm 0.012$\,d, is equal to one half of the orbital period $P_\text{orb} = 26.496$\,d. 
The fits of the polarization variability curves with Fourier series show a dominant contribution from the second harmonic which is typical for binary systems with circular orbits and nearly symmetric distribution of light scattering material with respect to the orbital plane. 
The continuous change of polarization with the orbital phase implies co-planarity of the orbit of the compact object and the Be star's decretion disk.
Using a model of Thomson scattering by a cloud that orbits the Be star, we obtained constraints on the orbital parameters, including a small eccentricity $e<0.2$ and periastron phase of $\phi_\text{p}\approx 0.6$, which coincides with the peaks in the radio, X-ray, and TeV emission. 
These constraints are independent of the assumption about the orientation of the decretion disk plane on the sky. 
We also extensively discuss the apparent inconsistency with the previous  measurements of the orbital parameters from radial velocities. 
By folding the photometry data acquired during a three-year time span with the orbital period, we found a linear phase shift of the moments of the brightness maximum, confirming the possible existence of superorbital variability. 
}

\keywords{binaries: general -- gamma rays: stars -- polarization -- stars: emission-line, Be -- stars: individual: \lsi}

\maketitle 
%

\section{Introduction}

Gamma-ray binaries constitute a subclass of high-mass binary systems with the emission peaking in the GeV band \citep[see recent review by][]{Chernyakova20arXiv}. 
In fact, \lsi\ is one of the best-studied $\gamma$-ray binaries and was observed in the last few decades over the whole range of the electromagnetic spectrum, from the radio to the very high-energy $\gamma$-rays \citep{Taylor, Paredes1994,Zamanov1999, Harrison, Abdo, Albert}. This binary system consists of a B0~Ve  star with a circumstellar disk \citep{Casares2005} and a compact companion star orbiting the primary star on an apparently eccentric ($e \geq 0.5$) orbit. The nature of the compact object, a black hole, or a neutron star is still unknown.  
 
Using a Bayesian analysis of 20 years of radio data,  \citet{Gregory2002} determined the orbital period $P_1 = 26.4960 \pm 0.0028$\,d. The compact object moving around the Be star and interacting with circumstellar matter produces the orbital variability seen in all parts of the spectrum \citep{Taylor1992, Mendelson, Leahy, Grundstrom2007}. A Lomb-Scargle timing analysis of 37 years of radio data resulted in the detection of the second period, $P_2 = 26.935 \pm 0.013$\,d \citep{MassiTorricelli}, which is consistent with a previously determined period of morphological changes in the radio structure, mapped in the VLBI images \citep{Massi}. Recently, the VLBA astrometry increased the accuracy of the period to $P_2 = 26.926 \pm 0.005$\,d \citep{Wu2018}. The beat of these two periods $P_{\text{beat}} = (P_1^{-1}-P_2^{-1})^{-1} \approx 1660$\,d is very close to the period of the long-term superorbital variability $P_{\text{sup}} \approx 1700$\,d observed in the radio \citep{MassiJaron}, H${\alpha}$ emission \citep{Zamanov2013}, X-rays \citep{Chernyakova2012,  Li2014}, and $\gamma$-rays \citep{Ackerman2013}.
    
 From the analysis of the optical spectropolarimetric data, \citet{Nagae2006, Nagae2009} showed that the polarization position angle (PA) was stable over the two-year period
 at $\theta\simeq 25{\degr}$. 
 According to these studies, the linear polarization in \lsi\ arises as a result of Thomson scattering in the Be star's decretion disk, and this disk is most likely co-aligned with the orbit of the compact companion \citep{Nagae2009}.
    
In this paper, we present the results of our \textit{BVR} high-precision photopolarimetric observational campaign of \lsi, which allowed us to determine the orbital period directly from the variability of the Stokes parameters of linear polarization for the first time and provide new constarints on the orbital parameters. 
In addition, we discuss our photometry data obtained for this object in terms of both orbital and superorbital variability.

\section{Photopolarimetric observations}

\subsection{Observed polarization}

The observations of \lsi\ were performed with the broad-band \textit{BVR} polarimeter Dipol-2 \citep{Piirola} mounted on the remotely controlled 2.2 m UH88 telescope at Mauna Kea Observatory and the 60 cm Tohoku telescope (T60) at Haleakala Observatory, Hawaii.
The object was observed during 140 nights from 2016 September 15 to 2019 September 1 (MJD 57646--58728).
Every night, 24 to 48 measurements of the normalized Stokes parameters $q$ and $u$ were made simultaneously in the \textit{B}, \textit{V}, and \textit{R}-passbands.
The total integration time with a typical ten-second exposure was 25--50 min. A summary of our observations is given in Table \ref{table:observations}. 
The observational errors of $q$ and $u$ were computed as the standard errors of the weighted mean values and are in the range of $0.01 - 0.03\%$.
In each of our five observing runs, the value of instrumental polarization was determined from observations of 15--20 nearby bright, unpolarized stars. 
The value of instrumental polarization is $\leq5 \times 10^{-5}$ for all passbands and is measured with the accuracy of a few parts per million ($10^{-6}$). To determine the PA zero point, the highly polarized standard stars HD 204827 and HD 25443 were observed. Detailed descriptions of the calibration and observation procedures can be found in \citet{Kosenkov} and \citet{Piirola2020}.  
Table \ref{table:polar} shows the observed average values of the polarization degree (PD) $P_\text{obs}$ and the PA $\theta_\text{obs}$.

 We also extracted the fluxes of \lsi\ and the closest field star (star 4, see Fig. \ref{fig:ISpolMap}) in the \textit{B}, \textit{V}, and \textit{R} passbands from our polarimetry images acquired with the T60 telescope, and we used them for the measurements of relative brightness variations of \lsi.

\begin{table}   
\caption{Log of polarimetric observations of \lsi.}             
\label{table:observations}      
\centering                          
\begin{tabular}{c c c c}        
\hline\hline                 
 Dates  & MJD & $N_\text{obs}$ & Telescope\\    
\hline                        

2016 Sep -- 2016 Oct & 57646 -- 57784 & 15 & UH88\\
2016 Oct -- 2017 Jan & 57646 -- 57784 & 20 & T60\\
2017 Sep -- 2018 Feb & 58026 -- 58158 & 50 & T60\\
2018 Aug -- 2019 Jan & 58337 -- 58474 & 32 & T60\\
2019 July -- 2019 Sep & 58684 -- 58727 & 23 &T60\\
\hline                                   
\end{tabular}
\end{table}

\begin{table*}   
\caption{Observed PD and PA of \lsi, interstellar polarization and average intrinsic polarization. 
}             
\label{table:polar}      
\centering                          
\begin{tabular}{c c c c c c c}        
\hline\hline                 
  & \multicolumn{2}{c}{Observed} & \multicolumn{2}{c}{Interstellar} & \multicolumn{2}{c}{Intrinsic}  \\   
Filter & $P_{\text{obs}}$ & $\theta_{\text{obs}}$ & $P_{\text{is}}$  & $\theta_{\text{is}}$ & $P_{\text{int}}$ & $\theta_{\text{int}}$  \\    
  &  (\%) &  (deg) &  (\%) &   (deg) & (\%) &   (deg) \\  
\hline                        
   \textit{B}  & $1.14\pm0.05$ & $139.5\pm1.5$ & $2.31 \pm 0.03$ & $117.3 \pm 0.5$ & $1.66 \pm 0.07$ &$13.5 \pm 0.9$  \\      
   \textit{V}  &  $1.21\pm0.05$ & $136.4\pm1.2$ & $2.40 \pm 0.03$ & $115.0 \pm 0.5$ & $1.72 \pm 0.05$ &$10.8 \pm 0.7 $ \\
   \textit{R}  & $1.25\pm0.04$ & $135.4\pm1.0$  &  $2.34 \pm 0.03$ & $113.9 \pm 0.5$ & $1.66 \pm 0.05$ &$8.4 \pm 0.7 $ \\ 
\hline                                   
\end{tabular}
\end{table*}

\subsection{Interstellar and intrinsic polarization}

\begin{figure}
   \centering
   \includegraphics[width=0.85\hsize]{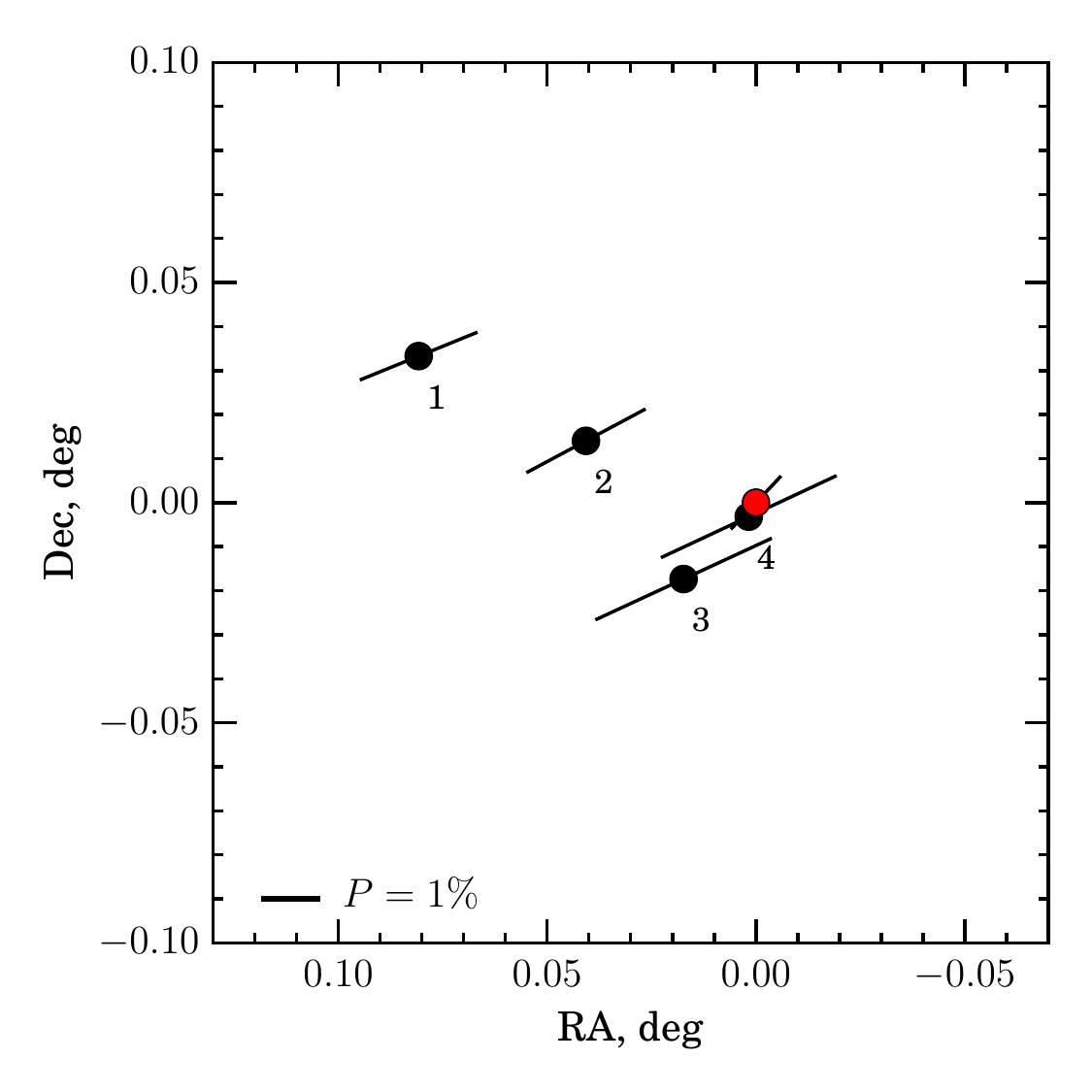}
      \caption{Polarization map of \lsi\ (red circle at the origin) and field stars (black circles) in the \textit{R} band. The length of the bars corresponds to the degree of linear polarization $P$, and the direction corresponds to the PA $\theta$ (measured from the north to the east). }
         \label{fig:ISpolMap}
\end{figure}

\begin{figure}
   \centering
   \includegraphics[width=0.90\hsize]{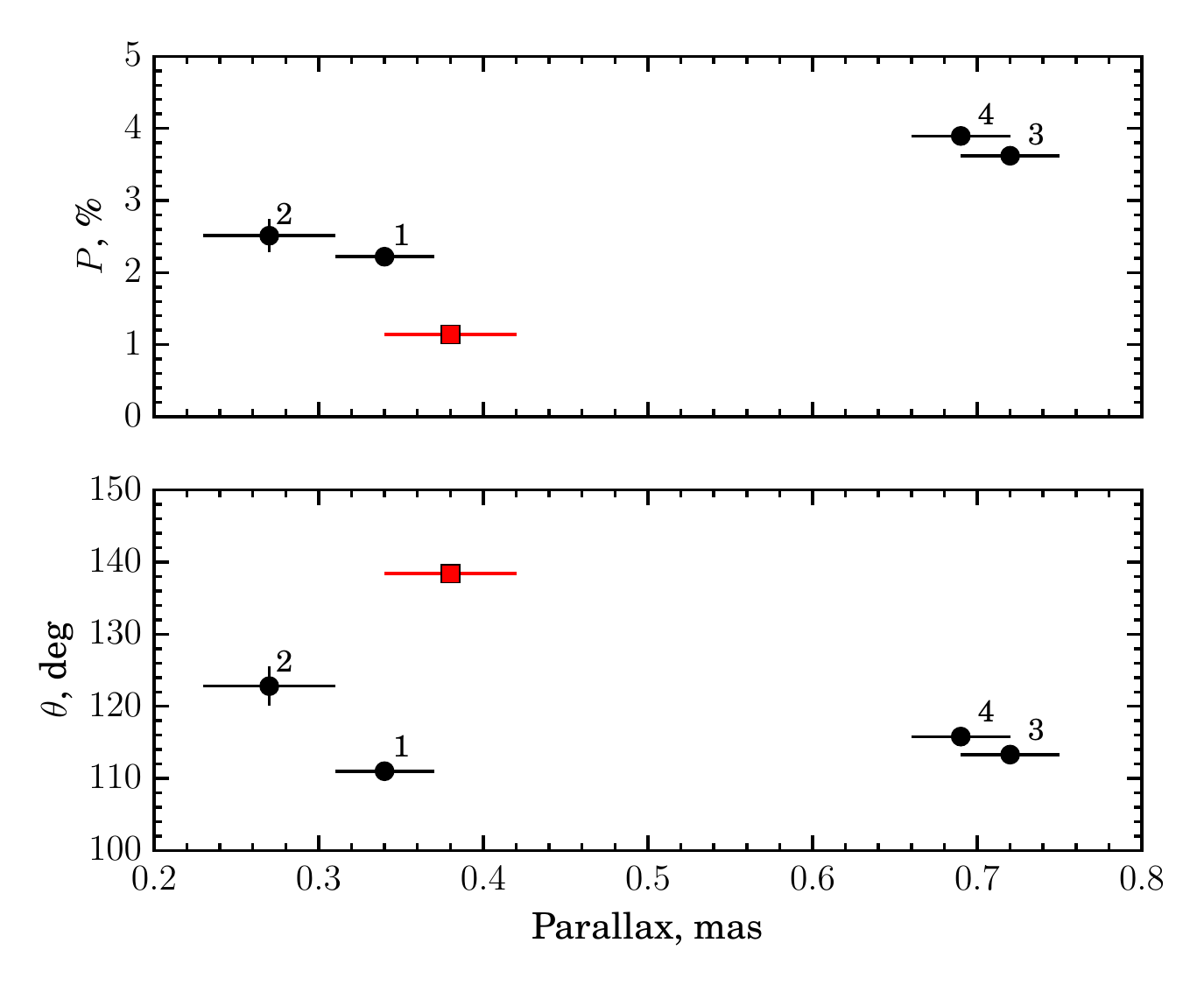}
      \caption{Dependence of observed PD $P$ and PA  $\theta$ on parallax (GaiaDR2) for \lsi\ (red) and field stars (black) in the \textit{B} band. The numbering of the field stars is the same as in Fig.\,\ref{fig:ISpolMap}.
       The error bars correspond to the $1\sigma$ errors. }
         \label{fig:ISpol}
\end{figure}

In order to obtain the degree and direction of the intrinsic linear polarization, $P_{\text{int}}$ and $\theta_{\text{int}}$, it is necessary to estimate the parameters of the interstellar (IS) polarization. For this purpose, we observed the polarization of four of the nearest field stars, located at the angular distance of $< 7 \arcmin$ from \lsi\ (Fig.\,\ref{fig:ISpolMap}) and with known parallaxes. Figure\,\ref{fig:ISpol} shows the dependence of the observed PD $P$ and PA $\theta$ on parallax for \lsi\ and the field stars (parallaxes are taken from the \textit{Gaia} Data Release 2; \citealt{GaiaDR2_1, GaiaDR2_2}).  The directions of the IS polarization for the field stars lie in the 110\degr-- 125\degr\ interval.
Assuming that the field stars are intrinsically unpolarized, we see that the degree of IS polarization decreases with distance (i.e., it grows with parallax) from $\approx$4\% to $\approx$2\% in the closest vicinity of the \lsi. This unusual behavior may result from the depolarization effect in several dust clouds with nearly orthogonal orientations of the Galactic magnetic field, located along the line of sight toward \lsi. 

\begin{figure}
   \centering
   \includegraphics[width=0.85\hsize]{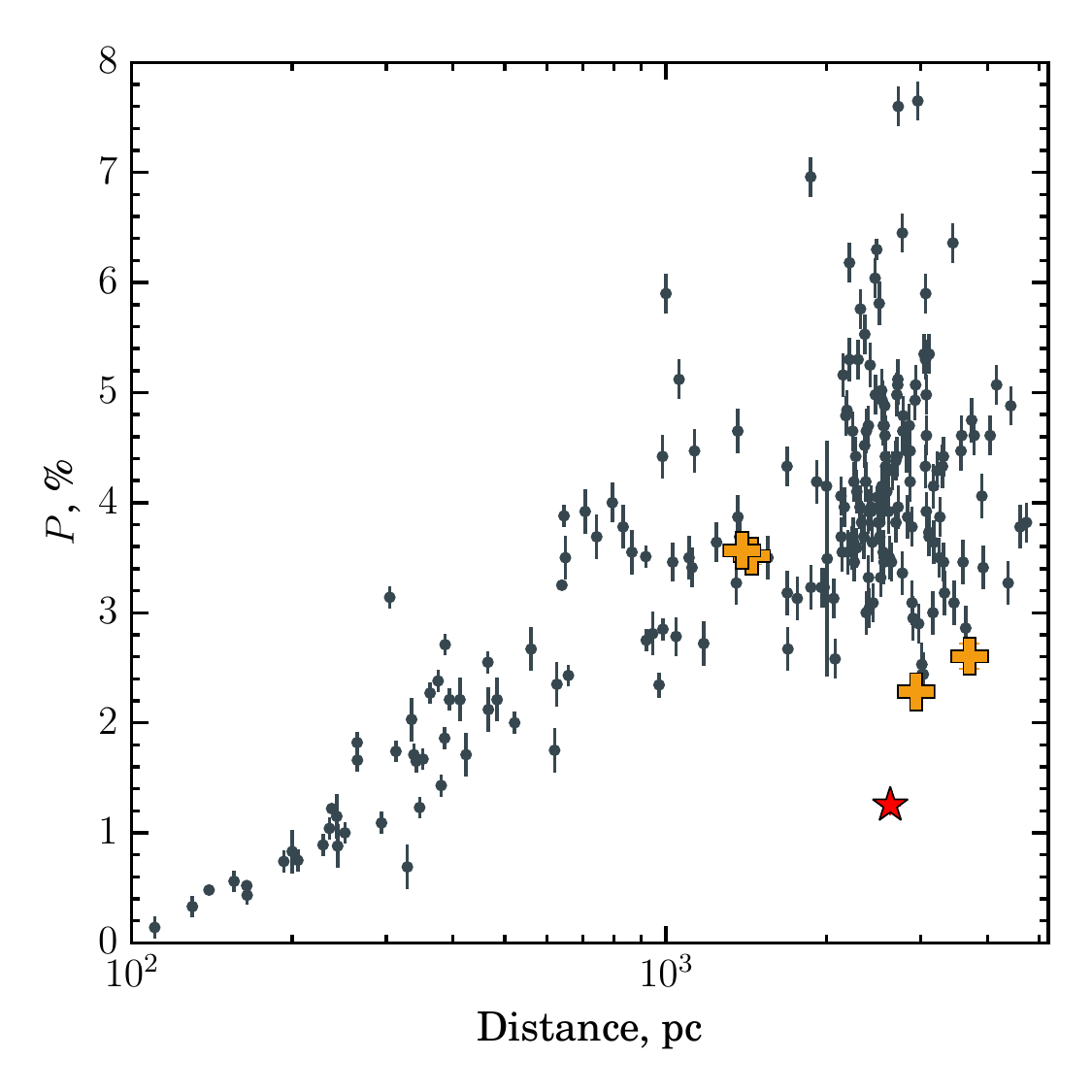}
      \caption{Dependence of linear PD $P$ \citep{Heiles2000} on distance (evaluated from the inverse of the GaiaDR2 parallaxes) for the stars in the $10\degr\times10\degr$ area around \lsi\  in the \textit{R} band. 
      The red star indicates our value of the observed polarization of \lsi, the orange crosses correspond to the four nearby field stars shown in Fig.\,\ref{fig:ISpolMap}. The vertical bar corresponds to the $1\sigma$ error in polarization.}
         \label{fig:polarParallax}
\end{figure}

For a more detailed study of IS polarization, we analyzed the data from Heiles' stellar polarization catalog \citep{Heiles2000} in a $10\degr \times 10\degr$ region around \lsi. The dependence of PD on distance for 232 stars in this area of the sky is shown in Fig.\,\ref{fig:polarParallax}. We see that there is a significant scatter of the IS PD for distant stars. Moreover, the PD of \lsi\ is smaller than that of all field stars at similar distances. It means that \lsi\ has a significant intrinsic polarization of which the direction does not match that of the IS polarization. The large scatter in the degree of IS polarization at the distances $d > 2$\,kpc may be linked to the complex structure of the IS medium inside the Heart Nebula (IC 1805) that is located in close proximity to \lsi. Thus, a careful approach, taking into account both proximity to the line of sight and proximity in distance, must be used for estimating the IS polarization for this object. 

We chose the average polarization of two nearby field stars (\#1 and \#2), which are close in distance to \lsi,\  as the best estimate for the IS polarization  in the direction of the binary. Table~\ref{table:polar} shows the estimated values $P_{\text{is}}$ and $\theta_{\text{is}}$ of the IS polarization in all passbands and the average values $P_{\text{int}}$ and  $\theta_{\text{int}}$ of the intrinsic polarization for \lsi, obtained after subtracting IS polarization. The values of IS polarization derived by us are in good agreement with those obtained by \citet{Nagae2006} from the polarization in H$\alpha$ emission line: $P_{\text{is}} = 2.20 \pm 0.18\%$, $\theta_{\text{is}} = 126\fdg 5 \pm 3\fdg7$. 

The values of the average intrinsic polarization of \lsi\  in the \textit{B}, \textit{V}, and \textit{R}-bands are the same within the errors. This suggests that Thomson scattering on free electrons in the disk around the Be star is likely the polarization mechanism responsible for the constant component of polarization in \lsi. This conclusion is in agreement with the results obtained by \citet{Nagae2006, Nagae2009}. It is interesting to note that the direction of the average intrinsic polarization in all passbands differs from the value of $25{\degr}$ derived by \citet{Nagae2006}. The PA of the intrinsic optical polarization of \lsi, obtained by us as the average for the \textit{BVR} bands, is $\simeq$11\degr.   
This difference, however, is most likely a result of uncertainty in the determination of the IS polarization and does not imply physical changes of the disk orientation with time. 
As we mentioned above, we used the average polarization of two (close in distance) field stars as an estimate of the IS polarization, whereas  \citet{Nagae2009} used polarization in H$\alpha$ and the empirical Serkowski law \citep{Serkowski73}, which is not applicable when the structure of the IS medium is  complex (i.e., consisting of multiple clouds of different properties), as is demonstrated by a peculiar  distance dependence of the PD for the nearby field stars shown in Fig. \ref{fig:polarParallax}.

The intrinsic polarization PA allows us to put constraints on the orientation of the  Be circumstellar disk on the sky. 
As was shown by \citet{Quirrenbach97} for four Be stars, the polarization vector is parallel to the projection of the rotation axis of the circumstellar disk. 
However, all their stars showed wavelength-dependent PD, which is an indication of the important role of bound-free hydrogen absorption in the envelope of a Be star. 
On the other hand, our intrinsic PD does not depend on the wavelength, implying that the electron scattering dominates. 
In this case, if the disk is optically thick, the polarization vector of radiation escaping from the disk may be perpendicular to the PA of the rotational axis \citep{ChaBre47,sob49,Wood1996}. 
In the following, we consider both possibilities.

\section{Orbital variability}
\label{sec:orb_var}

\subsection{Polarization variability}
\label{sec:pol_var}

\begin{figure}
   \centering
   \includegraphics[width=0.9\hsize]{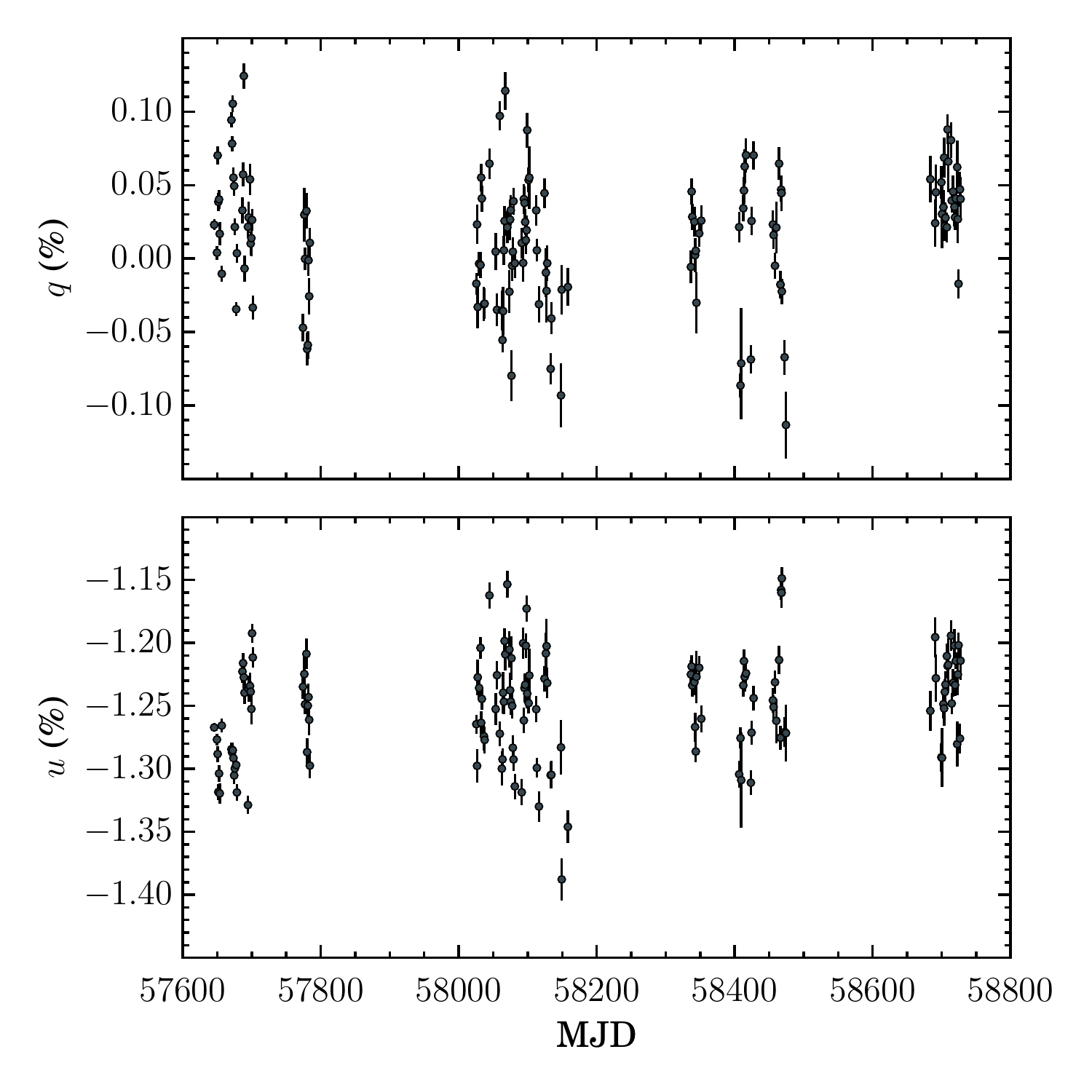}
      \caption{Observed Stokes parameters $q$ (\emph{top panel}) and $u$ (\emph{bottom panel}) with 1$\sigma$ errors for \lsi\  in the \textit{V} band, measured from 2016 September 15 to 2019 September 1 (MJD 57646--58728).  }
         \label{fig:FigStokesAll}
\end{figure}

\begin{figure}
        \centering
   \includegraphics[width=0.8\hsize]{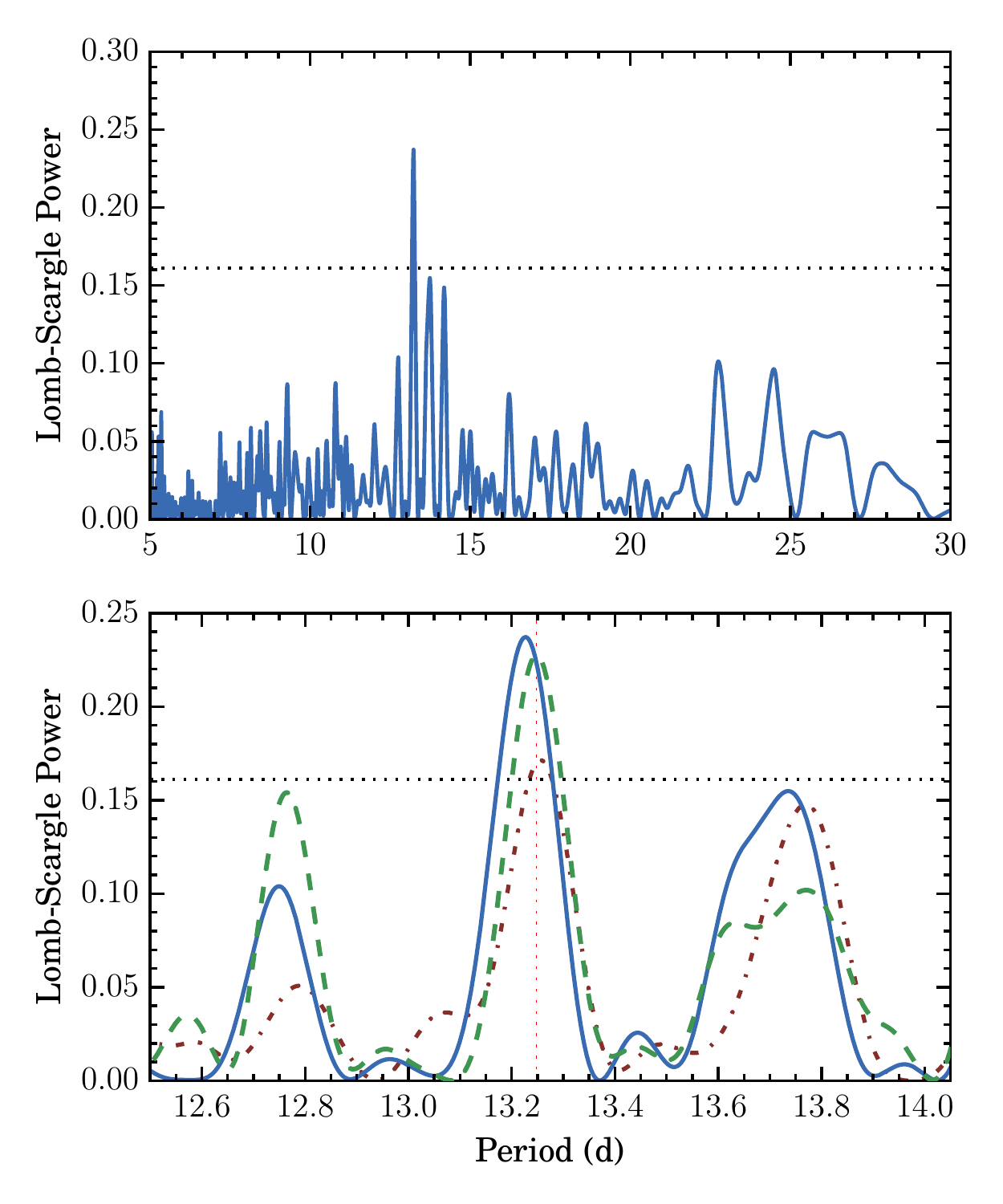}
      \caption{\emph{Top panel}: Lomb-Scargle periodogram for normalized Stokes parameter $u$ in the \textit{B} band. \emph{Bottom panel}: Lomb-Scargle periodogram zoom for the normalized Stokes parameter $u$ in the \textit{B}, \textit{V}, and \textit{R} bands (solid blue, dashed green, and dotted red curves, respectively). The horizontal dashed line corresponds to false alarm probability $\text{(FAP)} = 1\%$. The vertical red line corresponds to one half of the orbital period $P_\text{orb} = 26.496$\,d. } 
         \label{Fig:Scargle}
\end{figure}

The variability curves for the Stokes parameters $q$ and $u$ in the \textit{V} band, obtained over three years of observations of \lsi,  are shown in Fig. \ref{fig:FigStokesAll}. Our observations have revealed a small but significant variability with the amplitude of 0.2--0.3\% in all passbands. 
To study this variability, we performed a timing analysis of our \textit{BVR} polarimetric data. For this purpose, we applied the Lomb-Scargle method\footnote{\url{https://docs.astropy.org/en/stable/timeseries/lombscargle.html}}  of spectral analysis \citep{Lomb, Scargle}, using the {\sc AstroPy} package \citep{AstroPy}. 
The Lomb-Scargle periodograms for the normalized Stokes parameter $u$ in the \textit{BVR} passbands are shown in Fig. \ref{Fig:Scargle}. 
We see that the period of the highest peak in each band is close to one half of the orbital period $P_\text{orb} = 26.4960 \pm 0.0028$\,d, as is expected in binary systems. 
Despite significant nonperiodic scatter, this peak is clearly present in all passbands and both Stokes $q$ and $u$.    
For example, the period in the \textit{V} band in Stokes $u$ is $P_{V} = 13.25 \pm 0.06$\,d with a false alarm probability $\sim 10^{-5}$, which was independently estimated using  an analytical approximation of \citet{Baluev} and a bootstrap method from the {\sc AstroPy} package. The error on the found period was estimated in two different ways. In the first case, we simulated $10^4$ light curves, varying the observed data points around the mean values in accordance with the observational errors. In the second case, we also simulated $10^4$ light curves assuming sinusoidal variations with the fixed period $P = 13.25$~d and the amplitude, time stamps, and errors as in the observed data. For both cases, we computed the periodograms of the light curves and estimated an error on the found period as the half width of the  distribution of the highest peak positions. The value of the error in the first case was from three to ten times larger than in the second case, depending on the passband, due to the presence of the superorbital variability and other possible periodicities in the data. For the final error estimate, we chose the largest of the two.
The mean for the three passband periods in Stokes $u$ is $P_\text{pol} = 13.244 \pm 0.012$\,d,  which differs by less than 1$\sigma$ from $P_\text{orb}/2$. 
Thus, for the very first time, the orbital period of \lsi\  was obtained directly from the polarimetric measurements. 

\begin{figure}
   \centering
   \includegraphics[width=0.8\hsize]{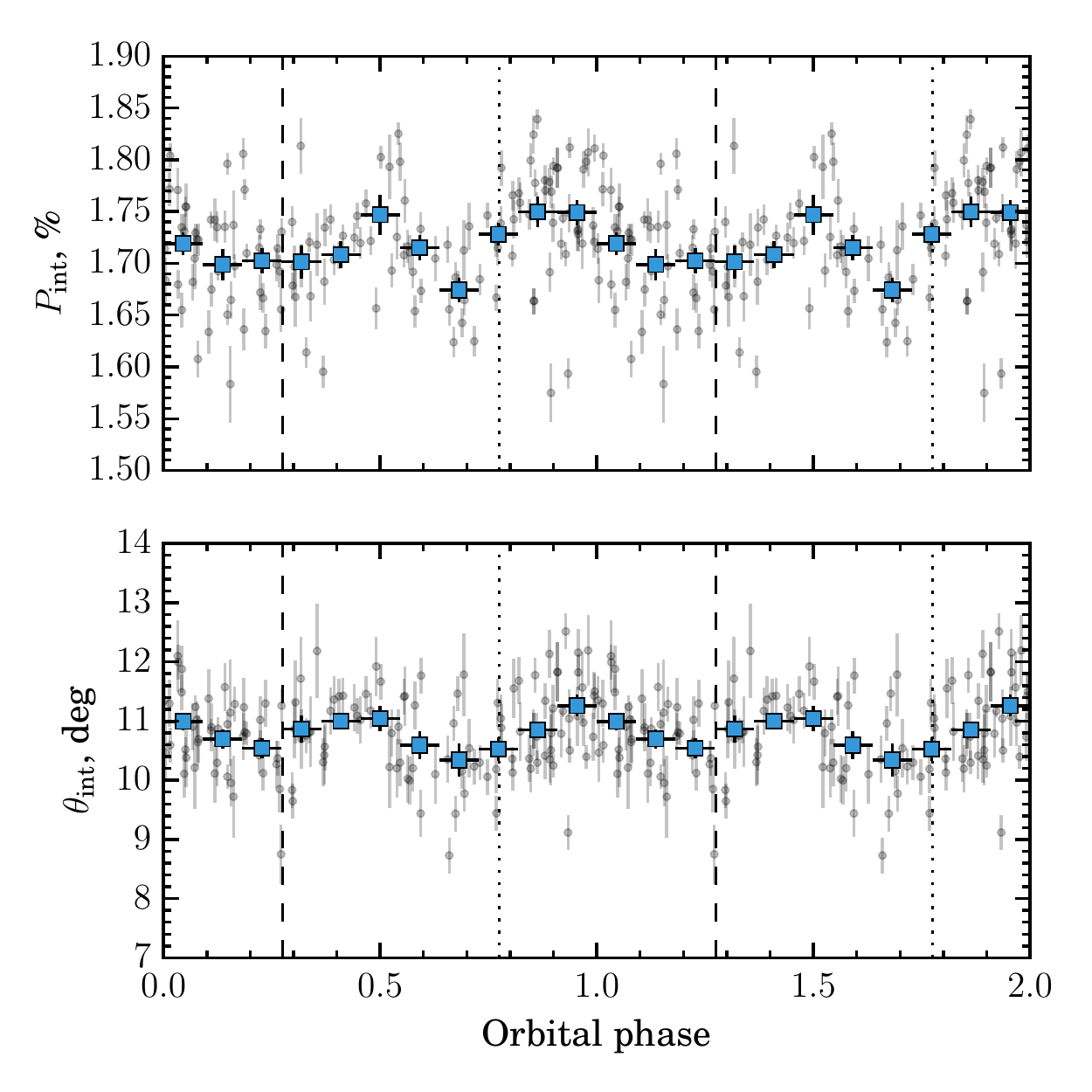}
      \caption{Orbital variability of intrinsic PD $P_\text{int}$ and PA $\theta_\text{int}$ of \lsi\ in \textit{V} band. The filled blue squares with 1$\sigma$ errors correspond to the average values of the individual observations (represented by gray circles with error bars) and the standard errors of the mean calculated within the phase bins of width $\Delta \phi = 0.091$. The vertical dashed lines correspond to the  phases of periastron ($\phi=0.275$) and apastron ($\phi=0.775$) as derived by \citet{Aragona2009}.} 
         \label{fig:intrinsic}
\end{figure}

\begin{figure*}
   \centering
   \includegraphics[width=0.8\linewidth]{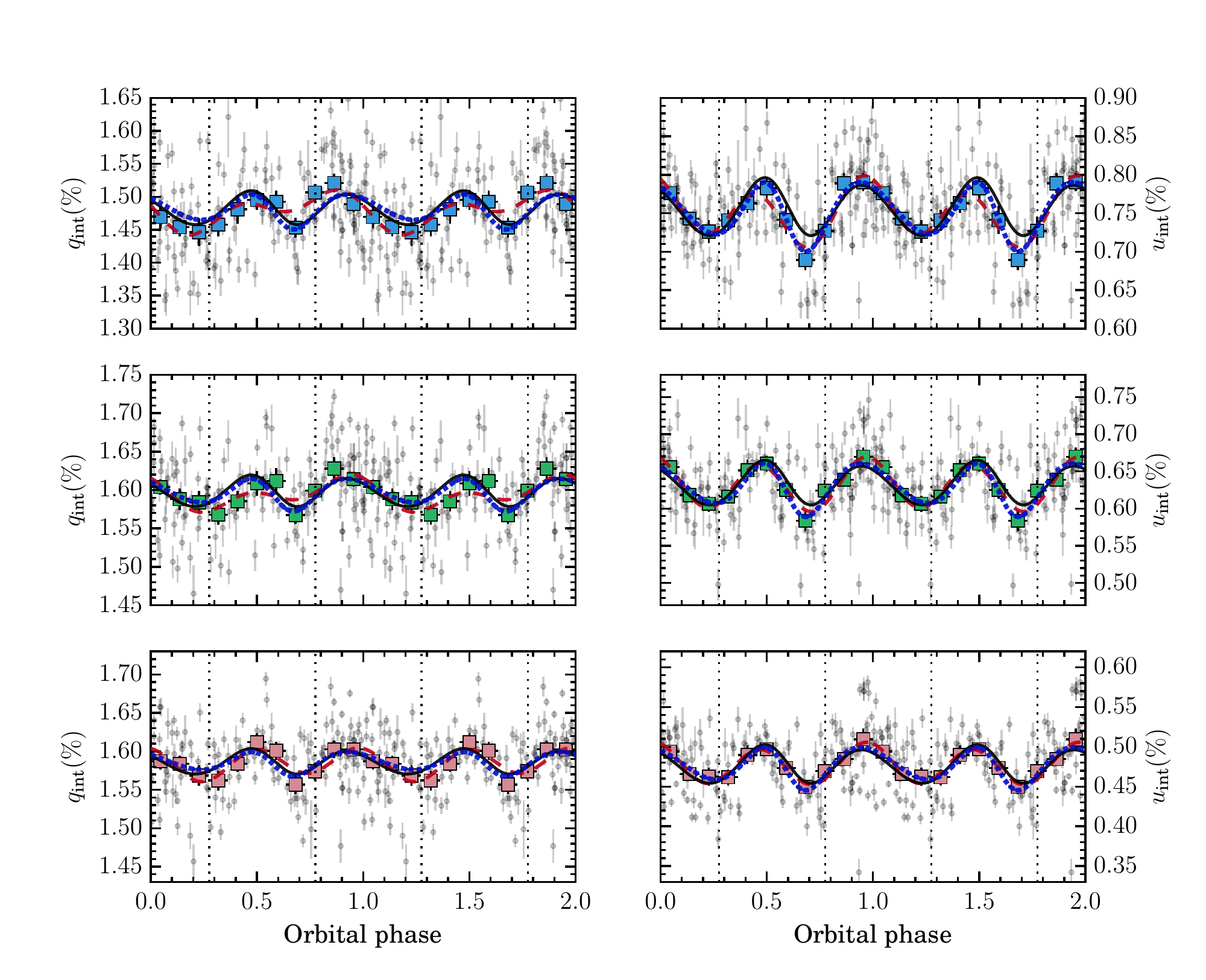}
      \caption{Variability of intrinsic Stokes parameters $q_\text{int}$ (\emph{left column}) and $u_\text{int}$ (\emph{right column}), in \textit{B}, \textit{V,} and \textit{R} bands (\emph{top, middle and bottom panels}, respectively) with the orbital phase folded with the period $P_\text{orb} = 26.496$\,d. The filled squares with 1$\sigma$ errors correspond to the average values of the individual observations (gray circles with error bars) and the standard errors of the mean calculated within phase bins of width $\Delta \phi = 0.091$. The vertical dashed lines correspond to the  phases of periastron ($\phi=0.275$) and apastron ($\phi=0.775$) as derived by \citet{Aragona2009}. The red dashed lines show the best fit with Fourier series given by Eq.\,(\ref{eq:Fourier}). 
      The black solid and blue dotted lines correspond to the best fit model of a scattering cloud on an eccentric orbit from Sect.\,\ref{sec:ecc_orbit}, for an $\Omega$ constrained close to $\theta_\text{int}$ and with a free $\Omega$, respectively. }
         \label{fig:Stokes}
\end{figure*}

Folding the data with the orbital period using the ephemeris given in \citet{Aragona2009}, we obtained the phase curves for the PD $P_\text{int}$ and the PA $\theta_\text{int}$ of the intrinsic polarization (Fig.\,\ref{fig:intrinsic}) and for the normalized Stokes parameters $q_\text{int}$ and $u_\text{int}$ (Fig.\,\ref{fig:Stokes}). 
The synchronization of the variable component of linear polarization in \lsi\  with the phase of the orbital period clearly indicates that it arises due to the orbital motion of the compact component. 
In the binary system with the compact companion and Be star, the extended hot region with enhanced electron density can be formed around or near a neutron star or a black hole in a process of their interaction with the disk material.
Thomson scattering of stellar photons on such a structure orbiting Be star can explain the regular small-amplitude variations in the observed polarization of \lsi.
Moreover, a continuous change of polarization implies co-planar orientation of the orbit and the Be star decretion disk. This conclusion directly follows on from the observed polarization variability and does not require any assumptions on the nature of compact companion \citep[see][]{Nagae2009}. 
There is also a strong nonperiodic component, which arises due to the long- and short-term changes in the distribution and density of the light scattering material in the Be decretion disk.

\begin{table*}
\caption{Fourier coefficients and their errors of the best fits to the data in \textit{BVR} bands with Eq.\,(\ref{eq:Fourier}).}             
\label{table:Fourier}      
\centering          
\begin{tabular}{c c c c c c c c c c c cc}     
\hline\hline       

Filter & $q_0$ & $u_0$ & $q_1$ & $u_1$ & $q_2$ & $u_2$ & $q_3$ & $u_3$ & $q_4$ & $u_4$ & $\chi^2$($q$)\tablefootmark{a}  & $\chi^2$($u$)\tablefootmark{b}  \\ 
\hline

   \multirow{2}{*}{\textit{B}}  & $1.479$ & $0.751$ & $-0.0005$ & $0.014$ & $-0.021$ & $0.005$ & $0.008$ & $0.029$ & $-0.017$ & $-0.021$ & 
   7.1 & 6.3 \\ 
   
   & $\pm0.005$ & $\pm0.004$ & $\pm0.007$ & $\pm0.005$ & $\pm0.007$ & $\pm0.006$ & $\pm0.007$ & $\pm0.006$ & $\pm0.007$ & $\pm0.005$& & \\
   \multirow{2}{*}{\textit{V}}  & $1.594$ & $0.631$ & $0.009$ & $0.006$ & $-0.011$ & $0.001$ & $0.012$ & $0.030$ & $-0.007$ & $-0.013$ &  7.9 & 3.7  \\
   & $\pm0.004$ & $\pm0.003$ & $\pm0.006$ & $\pm0.005$ & $\pm0.005$ & $\pm0.005$ & $\pm0.005$ & $\pm0.005$ & $\pm0.006$ & $\pm0.004$& & \\

   \multirow{2}{*}{\textit{R}} & $1.585$ & $0.477$ & $0.0006$ & $0.005$ & $-0.006$ & $-0.0002$ & $0.019$ & $0.022$ & $-0.002$ & $-0.009$ &  7.6 &  2.2 \\
   & $\pm0.003$ & $\pm0.003$ & $\pm0.004$ & $\pm0.004$ & $\pm0.005$ & $\pm0.004$ & $\pm0.005$ & $\pm0.004$ & $\pm0.005$ & $\pm0.004$& & \\
\hline                  
\end{tabular}
\tablefoot{
\tablefoottext{a}{$\chi^2$ values for Stokes $q$ fit with 6 dof. }
\tablefoottext{b}{$\chi^2$ values for Stokes $u$ fit with 6 dof. }
}
\end{table*}

The dominant second harmonic in the variability of the Stokes parameters is typical for a binary system with symmetric density distribution of the scattering matter about the orbital plane and a circular orbit \citep{Brown1978}. The polarization arising from light scattering on gaseous structures like clouds, streams, and so on, peaks at orbital phases where the scattering angle is close to $90\degr$. For a co-rotating envelope and a circular binary orbit, this gives two prominent symmetric peaks in polarization separated by the phase interval $\simeq0.5$. In the case of a (weak) asymmetry of the distribution of the light scattering material about the orbital plane, the first harmonic appears. This is apparently observed in \lsi. As we see in Figs. \ref{fig:intrinsic} and \ref{fig:Stokes}, there are two nearly symmetric peaks near the phases 0.5 and 1.0. They are most pronounced in the variations of the Stokes $u$ and the PA $\theta$.

\subsection{Modeling polarization with circular orbit}
\label{sec:circular}

According to the common approach \citep{Brown1978}, the phase curves of the Stokes parameters in binary systems with a circular orbit and co-rotating light scattering envelope can be represented through a Fourier series of the orbital longitude $\lambda = 2\pi\phi$ (where $\phi$ is a phase of the orbital period):
\begin{equation}\label{eq:Fourier}
\begin{aligned}
q_\text{int}  & = q_0 + q_1\cos{\lambda} + q_2\sin{\lambda} + q_3\cos{2\lambda} + q_4\sin{2\lambda}, \\
u_\text{int}  & = u_0 + u_1\cos{\lambda} + u_2\sin{\lambda} + u_3\cos{2\lambda} + u_4\sin{2\lambda} .
\end{aligned}
\end{equation}

\begin{table*}
    \centering
    \caption{Best fit parameter estimates for the model described in Sect.\,\ref{sec:ecc_orbit} and Appendix \ref{sec:appendix}. Errors are 1$\sigma$. 
    }
    \label{tab:tbl_ecc_fit}
    \begin{tabular}{cccccccccc}
    \hline
    \hline
     Filter  & $e$\tablefootmark{a} & $i$\tablefootmark{a} & $\Omega$\tablefootmark{a} & $\lambda_\mathrm{p}$\tablefootmark{a} & $\phi_\mathrm{p}$\tablefootmark{a} & $q_0$ & $u_0$ & $f_0$ & $\chi^2$/dof\\
     & & (deg) & (deg) & (deg) &   & (\%) & (\%) & (\%) &  \\
     \hline
    \multicolumn{10}{c}{$\Omega = \theta_\text{int}\pm 10\degr$} \\
     \textit{B} & \multirow{3}{*}{0.06 $\pm$ 0.02} & \multirow{3}{*}{86 $\pm$ 3} & \multirow{3}{*}{28 $\pm$ 3} & \multirow{3}{*}{$146 \pm 22$} & \multirow{3}{*}{0.62 $\pm$ 0.07} & 1.458 $\pm$ 0.007 & 0.721  $\pm$ 0.007 & 0.23  $\pm$ 0.03 & \multirow{3}{*}{67/52} \\
     \textit{V} & & & & & & 1.579 $\pm$ 0.005 & 0.605  $\pm$ 0.006 & 0.18 $\pm$ 0.03 & \\
     \textit{R} & & & & & & 1.571 $\pm$ 0.005 & 0.454  $\pm$ 0.005 & 0.15 $\pm$ 0.02 & \\
    \hline
    \multicolumn{10}{c}{Free $\Omega$} \\
    \textit{B} & \multirow{3}{*}{0.11 $\pm$ 0.03} & \multirow{3}{*}{87 $\pm$ 3} & \multirow{3}{*}{120 $\pm$ 2} & \multirow{3}{*}{225 $\pm$ 13} & \multirow{3}{*}{0.59 $\pm$ 0.04} & 1.503 $\pm$ 0.006 & 0.790  $\pm$ 0.006 & 0.24  $\pm$ 0.03 & \multirow{3}{*}{50/52} \\
    \textit{V} & & & & & & 1.614 $\pm$ 0.005 & 0.661  $\pm$ 0.005 & 0.19 $\pm$ 0.03 & \\
    \textit{R} & & & & & & 1.599 $\pm$ 0.004 & 0.499  $\pm$ 0.004 & 0.14 $\pm$ 0.02 & \\
    \hline
    \end{tabular}
    \tablefoot{
\tablefoottext{a}{Parameter is the same for all three filters.} 
}
\end{table*}

\begin{figure*}
\centering
\includegraphics[width=0.95\linewidth]{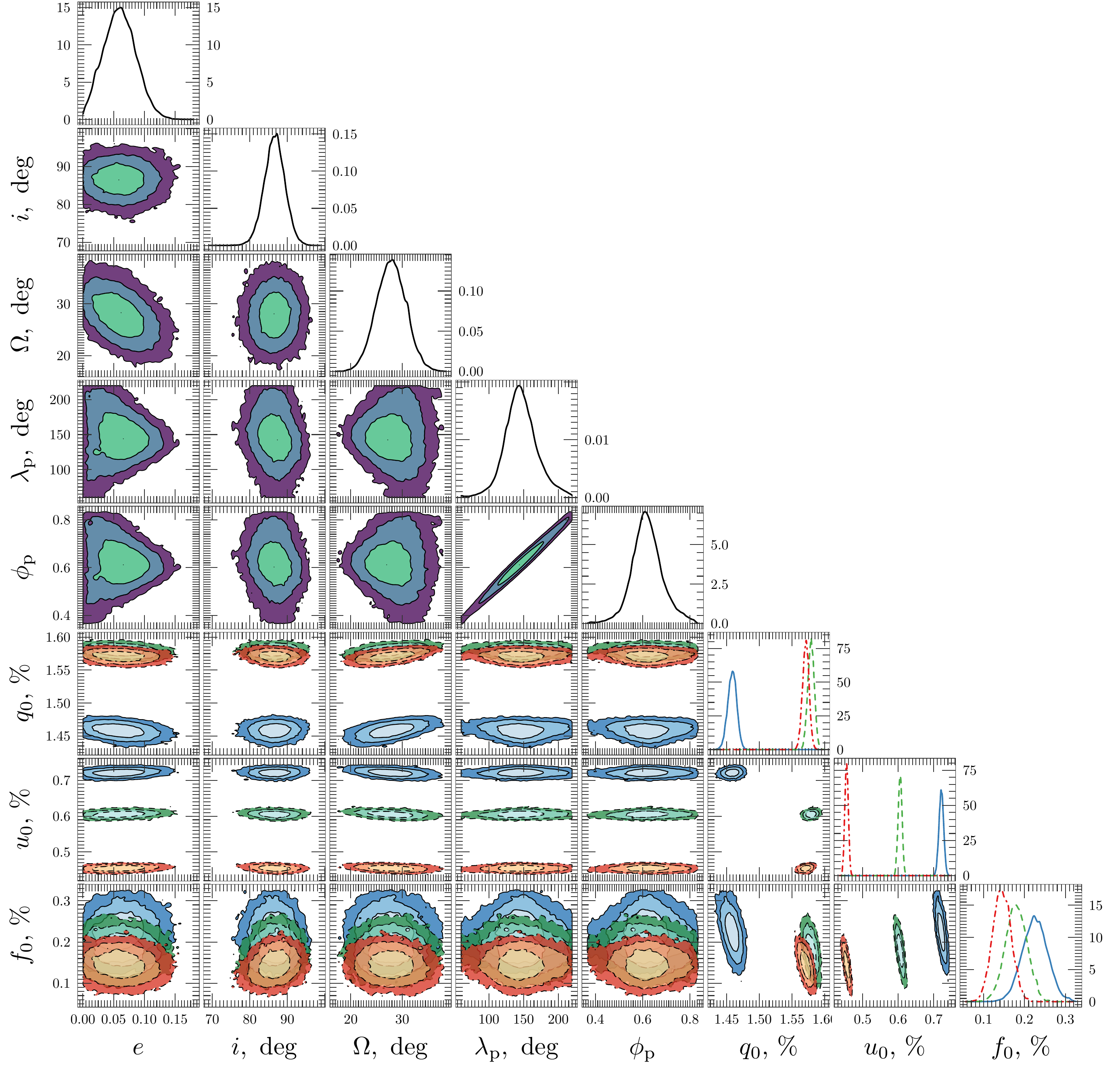}
\caption{Posterior distributions for parameters of model  described in Sect.\,\ref{sec:ecc_orbit} and Appendix\,\ref{sec:appendix} for $\Omega=  \theta_\text{int} \pm 10\degr$. 
\emph{Diagonal panels}: distributions of model parameters. The blue solid, green dashed, and red dot-dashed lines correspond to per-passband \textit{B}, \textit{V}, and \textit{R} distributions, respectively.
\emph{Lower-triangle panels}: joint posterior distributions of two parameters. The green, blue, and violet contours correspond to 0.68, 0.95, and 0.997 probability levels of parameters $i$, $\Omega$, $\lambda_\text{p}$, and $\phi_\text{p}$. The shades of blue, green, and red correspond to the same probability levels for \textit{B}, \textit{V}, and \textit{R} filters, respectively. } 
    \label{fig:posteriors}
\end{figure*}

\begin{figure*}
  \centering
\includegraphics[width=0.40\linewidth]{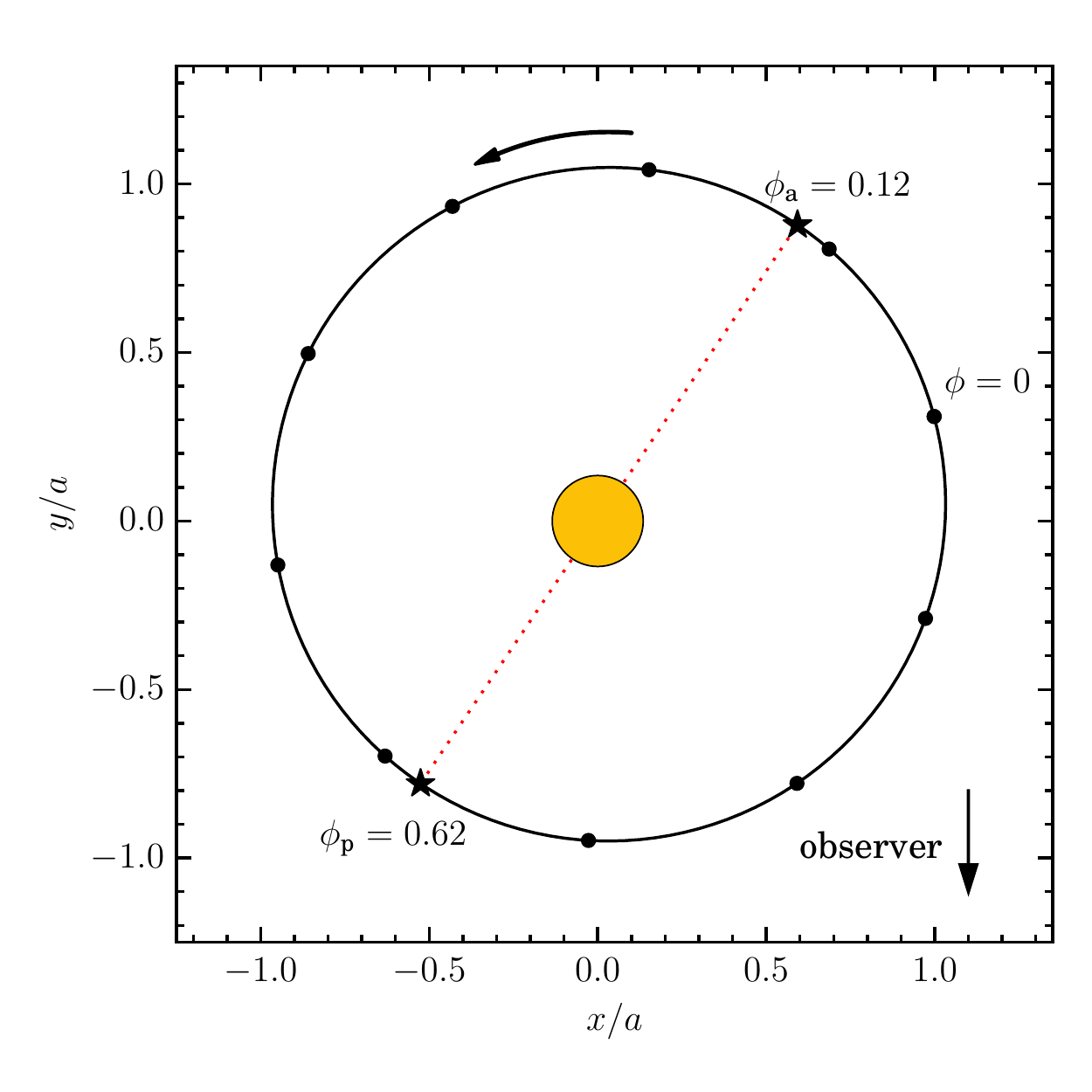}  
\hspace*{1cm}
\includegraphics[width=0.40\linewidth]{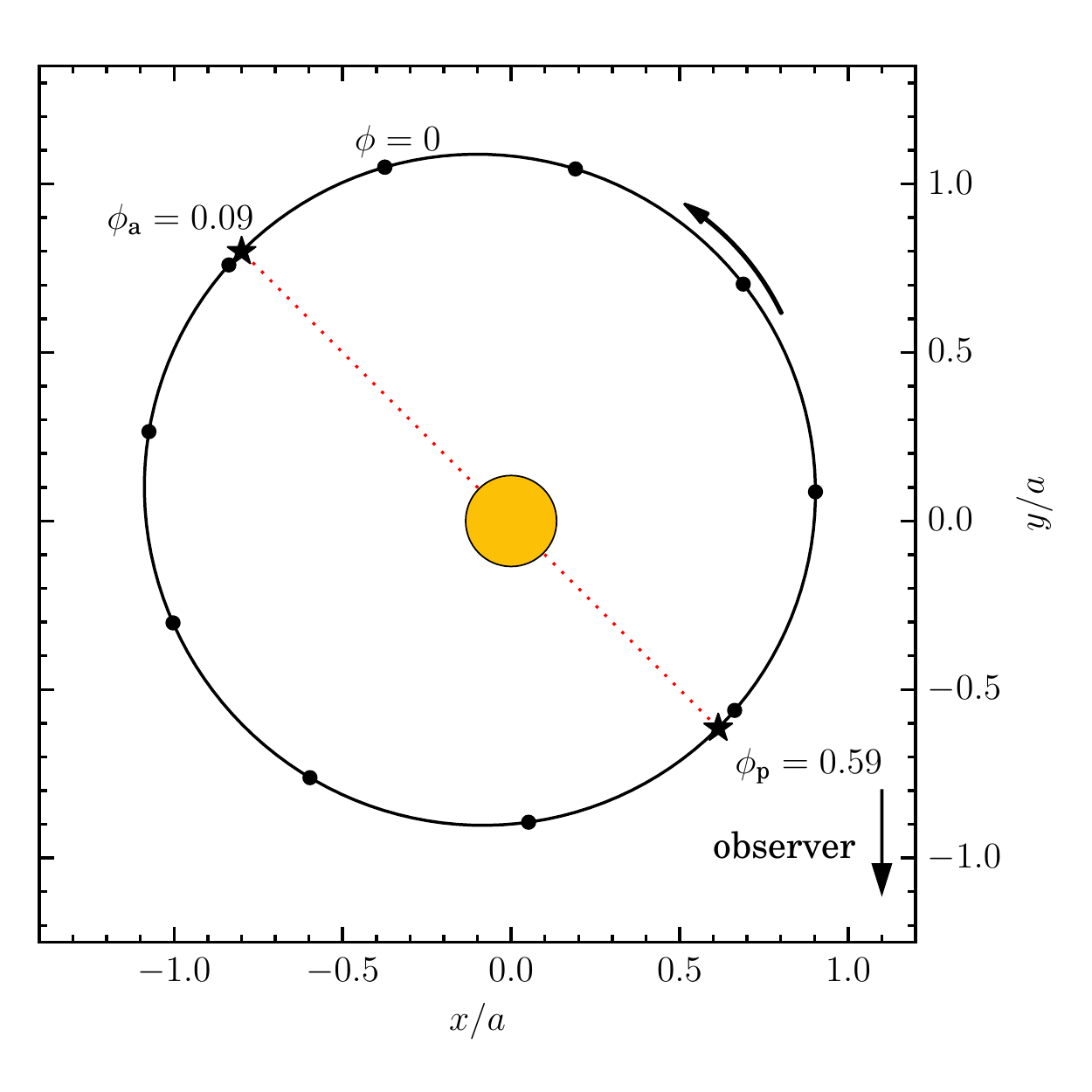}  
      \caption{Relative orbit of a compact object in \lsi\ around Be star (yellow circle at the origin), which lies at the ellipse focus. The orbital parameters are taken from Table \ref{tab:tbl_ecc_fit}. The red dashed line is the major axis of the orbit. The black dots on the ellipse are spaced by the $\Delta \phi = 0.1$. 
      Because of a 180\degr\ degeneracy in $\lambda_\text{p}$, the orbit can be also rotated by 180\degr.
      \textit{Left panel}: orbit for constrained orientation relative to the decretion disk plane with $\Omega=  \theta_\text{int} \pm 10\degr$. 
      \textit{Right panel}: orbit assuming free $\Omega$. 
 }
\label{fig:orbital_elems}
\end{figure*}

We fit the intrinsic Stokes parameters $q_\text{int}$ and $u_\text{int}$ with these functions, using  $q_0$--$q_4$ and $u_0$--$u_4$ as free parameters. 
The best fit parameters together with the $\chi^2$ values and the degrees of freedom (dof) are given in Table\,\ref{table:Fourier}. 
The best fit curves are shown in Fig.\,\ref{fig:Stokes} (red dashed lines). We see that variabilities of Stokes parameters $q$ and $u$ are dominated by the second harmonic of the orbital period. This is expected, because of the position of the highest peak in the Lomb-Scargle periodogram close to the frequency corresponding to one half of the orbital period. The parameters 
\begin{equation}
A_q= \sqrt{\frac{q^2_4+q^2_3}{q^2_2+q^2_1}},\quad A_u = \sqrt{\frac{u^2_4+u^2_3}{u^2_2+u^2_1}}  ,
\end{equation}
giving the ratio of the amplitudes of the second  to the first harmonic, attain the values $A_q = 0.9, 1.0, 3.3$ and $A_u = 2.5, 5.5, 5.4$ for the \textit{B},  \textit{V}, and  \textit{R} bands, respectively.

It is possible to derive the inclination $i$ of the binary orbit and the  position angle $\Omega$ of the projection of the orbital axis (see Eqs.\,\ref{eq:q_rotate}, \ref{eq:u_rotate}) from the best fit Fourier coefficients in Eq.\,(\ref{eq:Fourier}) to the observed (or intrinsic) Stokes parameters $q$ and $u$ \citep{Drissen}:
\begin{equation}\label{eq:Drissen_i}
\left(\frac{1 - \cos{i}}{1 + \cos{i}}\right)^4 = \frac{(u_3 + q_4)^2 + (u_4 - q_3)^2}{(u_4 + q_3)^2 + (u_3 - q_4)^2}
\end{equation}
\begin{equation}\label{eq:Drissen_omega}
\tan{2\Omega} = \frac{A + B}{C + D},\\
\end{equation}
where
\begin{equation}
\begin{aligned}
A &= \frac{u_4 - q_3}{(1 -\cos{i})^2},\quad B = \frac{u_4 + q_3}{(1 +\cos{i})^2} ,\\
C &= \frac{q_4 - u_3}{(1 +\cos{i})^2},\quad D = \frac{u_3 + q_4}{(1 -\cos{i})^2}. 
\end{aligned}
\end{equation}

However, there is a bias in the polarimetrically derived  orbit  inclination because of  the  unavoidable noise (finite accuracy) in the polarization data \citep{Aspin1981, Simmons1982, Wolinski1994}. The inclination of orbit $i$ derived from the best fit Fourier coefficients is always biased toward a higher value. For noisy data, the inclination approaches $90{\degr}$ with wide confidence intervals extending to very low values \citep[see][]{Wolinski1994}. Similar bias can also be induced by stochastic noise, arising from an intrinsic nonperiodic component of the polarization variability \citep{Manset2000}. As the errors on $q$ and $u$ increase, a straight line becomes an acceptable fit to the ($q$, $u$) light curves. Because a straight-line fit to the ($q$, $u$) data yields a $90{\degr}$ inclination for the system \citep{Brown1978}, the derived value of $i$ is biased toward $90{\degr}$ when a high noise level is present in the data.

Due to the strong nonperiodic component in the polarization variability of \lsi, the inclination of the orbit derived from the Fourier coefficients for all passbands is close to $90{\degr}$, with $2\sigma$ confidence intervals extending down to values of $i\leq20{\degr}$. Thus, assuming a (nearly) circular orbit in \lsi, we can only say that our polarization data are consistent with orbit inclination $i$ in the range $\approx 20{\degr} - 90{\degr}$. 
The position angle $\Omega$, derived from Eq.\,\eqref{eq:Drissen_omega}, is about 30\degr\ for all passbands.  
As was shown by \citet{Wolinski1994}, there is no bias in the value of $\Omega$ derived from polarimetry. This is due to the fact that the value of $\Omega$ is determined by the positions of the polarization maxima and minima, which are not affected by the noise in the same way as the amplitude and shape of the polarization variability curve.

\subsection{Modeling polarization with an eccentric orbit}
\label{sec:ecc_orbit}

The latest value of eccentricity $e$ derived for \lsi\  from the radial velocity variations in the optical spectral lines is $0.54 \pm 0.03$ \citep{Aragona2009}. For such an eccentric orbit, one should expect fast changes in polarization occurring near the periastron passage \citep{Brown1982, Simmons1984}. According to the orbital geometry shown in Fig. 3 of \citet{Aragona2009}, we should expect one sharp peak just before, and another one soon after, the periastron passage, which occurs at phase 0.275. These two peaks must be followed by a gradual change of polarization during the remaining part of the orbital motion toward apastron. Such a behavior of intrinsic polarization has been detected in the interacting binary HD 187399, which has an orbit eccentricity of $e \simeq 0.4$ \citep{Berdyugin1998}, but this is not observed in \lsi.

Using the approach proposed by \citet{Simmons1984}, we tried to model the observed shape of the polarization variability curve in \lsi. We considered a simple scenario of a small cloud of free electrons orbiting the central illuminating source and scattering its radiation (see Appendix~\ref{sec:appendix} for the model). Under this assumption, the modeled Stokes parameters depend on the eccentricity $e$, the orbit inclination $i$, as well as on the longitude of the periastron $\lambda_\text{p,}$ and typical scattering fraction $f_0$ (see Eq.\,\ref{eq:model_Stokes}). Polarization, produced by such a system, depends on time through the orbital longitude $\lambda$, which can be connected to the observed orbital phase $\phi$ through an eccentric anomaly (Eq.\,\ref{eq:eccent_anom}) and a Kepler equation (Eq.\,\ref{eq:meananom}).  
The orbital phase is then corrected for the phase of the periastron $\phi_\text{p}$ (corresponding to $\lambda = \lambda_\text{p}$) using Eq.\,(\ref{eq:phiorb}). 
The model Stokes pseudo-vector should also be rotated to account for the  position angle $\Omega$ of the projection of the orbital axis (see Eqs.\,\ref{eq:q_rotate} and \ref{eq:u_rotate}) relative to the north. 
The constant levels of intrinsic polarization corresponding to the emission from the decretion disk is represented by the pseudo-vectors $(q_0, u_0)$.

\begin{figure*}
\centering
\includegraphics[width=0.95\linewidth]{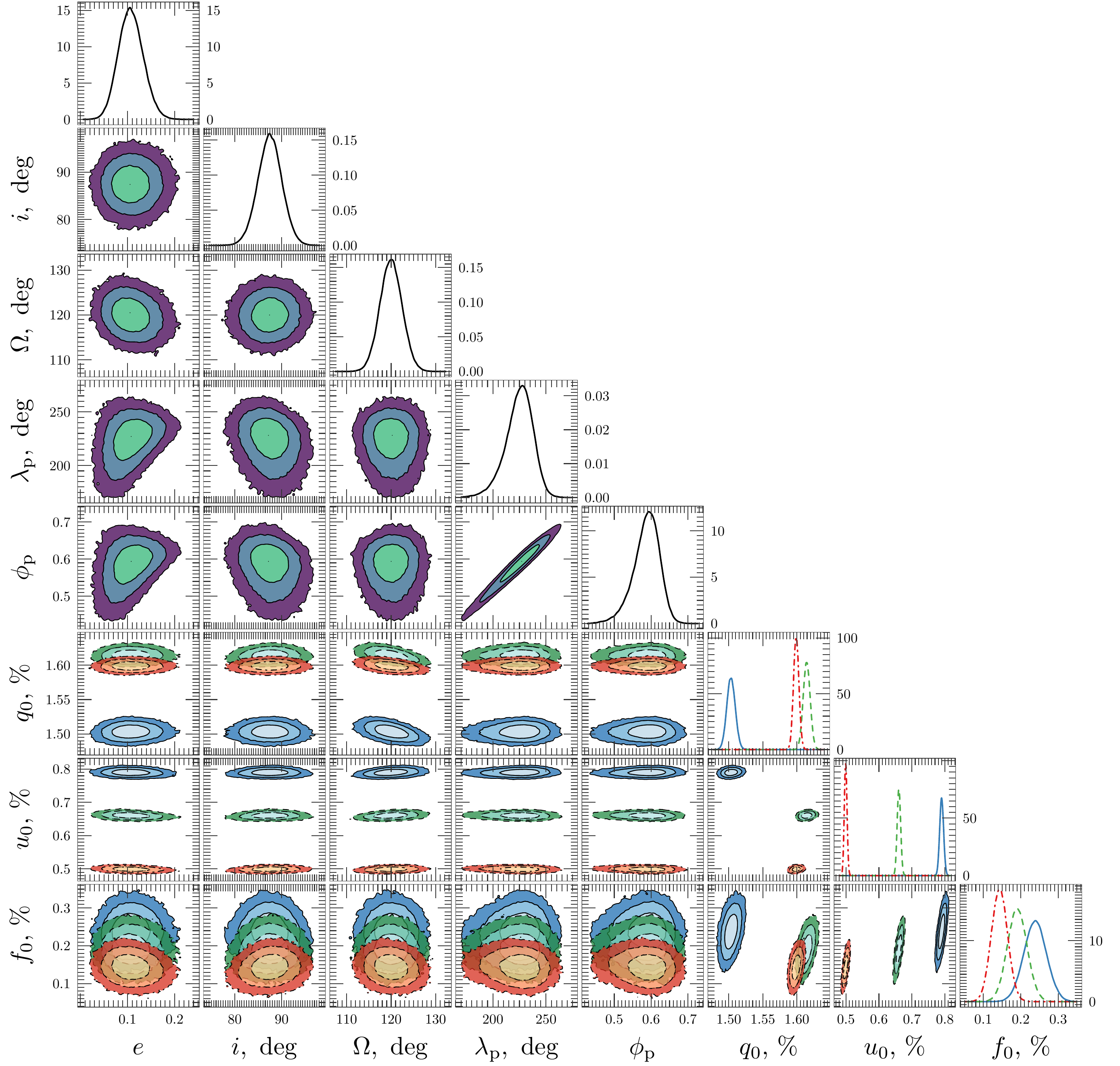}
\caption{Same as Fig.\,\ref{fig:posteriors}, but for free  $\Omega$.    } 
    \label{fig:posteriors_free}
\end{figure*}

We employed Bayesian inference to fit this model to the intrinsic polarization light curves of \lsi\ in three filters simultaneously, allowing different levels of constant polarization $(q_0, u_0)$ and a typical scattering fraction $f_0$ in each band.
We applied a Hamiltonian Monte Carlo algorithm \citep{Duane1987} implemented in the {\sc greta} package \citep{greta2019} for {\sc R} \citep{rstats}, which utilizes the {\sc tensorflow} backend (see \citealt{tensorflow2015}). 
First, we took the eccentricity $e=0.54$ and the argument of the periastron $\omega=\lambda_\text{p} \pm 90^{\circ}=40{\degr}$ from \citet{Aragona2009}, but other parameters were allowed to vary in the broadest possible intervals.   
We were not able to fit the observed variability of the Stokes parameters $q$ and $u$ for the whole range of reasonable inclinations (i.e., $i = 20\degr - 90\degr$) resulting in $\chi^2/\mbox{dof}>180/54$. Then, we fixed only eccentricity $e \simeq 0.5$ and tried to fit the curves by varying other orbital parameters. We could only fit variations of one Stokes parameter reasonably well, with the resulting $\chi^2/\mbox{dof} = 177/53$. 
Thus, we were not able to find a set of orbital parameters providing a sufficiently good fit for both Stokes $q$ and $u$ for $e\geq 0.5$. This can be simply explained by the influence of eccentricity on the shape of the polarization curve. We can see that in our data (Fig. \ref{fig:Stokes}) the distance between the neighboring maxima is  $\Delta \phi \approx 0.4$, while in the case of the eccentricity $e \approx 0.5$, it should be $\Delta \phi \approx 0.15$.

\begin{figure*}
\centering
\includegraphics[width=0.80\linewidth]{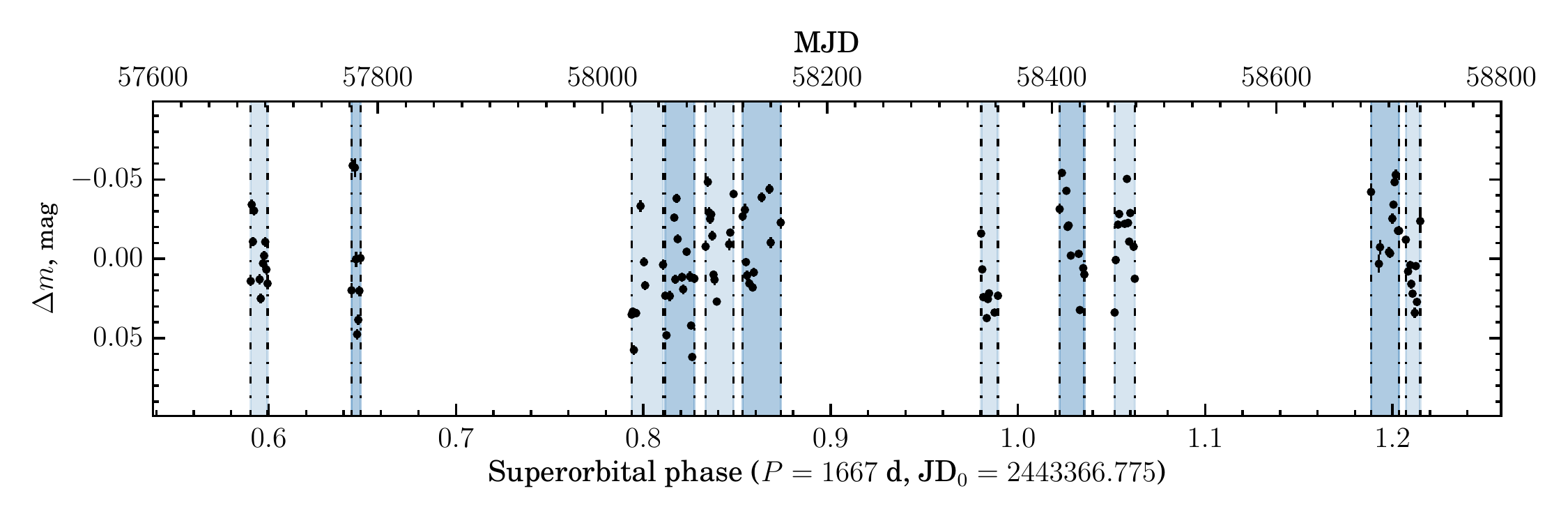}
\caption{Variation of relative amplitude of \lsi\ around the mean in the \textit{V} passband as a function of superorbital phase. 
The shaded blue bands correspond to the eleven observing seasons (S1--S11). } 
         \label{Fig:Timeline}
\end{figure*}

Finally, we allowed all parameters, including eccentricity $e$, to vary.
At the first step, we assumed that the projection of the decretion disk axis on the sky coincides with the PA of the average intrinsic polarization. 
If the orbit of the compact object and the decretion disk are nearly co-planar (see Sect.\,\ref{sec:pol_var}), we expect that $\Omega\approx\theta_\text{int}$.  
We thus limited the prior distribution of $\Omega$ to follow a Gaussian with the peak at 11\degr\ ($\theta_\text{int}$ from Table~\ref{table:polar}) with an arbitrarily chosen standard deviation of 10\degr\ and truncation interval of $[-10\degr; 40\degr]$.
The best fit parameters together with the values of $\chi^2$ and the dof are given in the upper part of Table~\ref{tab:tbl_ecc_fit}, and a comparison of the model with the data is shown in Fig.\,\ref{fig:Stokes}.
The posterior distributions are shown in Fig.~\ref{fig:posteriors}. 
A corresponding orbital geometry is demonstrated in the left panel of Fig.~\ref{fig:orbital_elems}.

We see that our eccentric orbit fit gives an upper limit on $e$ of 0.15, with the best fit value $e \simeq 0.06$.
The high value of inclination $i$ obtained from the eccentric orbit fit is most likely a result of a similar bias as for the circular orbit model.
The best fit value of $\Omega\approx 28\degr$ coincides with the value of 30\degr\ obtained from the fits with the circular orbit model in Sect.\,\ref{sec:circular}. The best fit phase of the periastron $\phi_\text{p}\approx0.62$ differs significantly from the usually assumed value of 0.275. 
We note that some of the parameters are degenerate: for example, $\lambda_\text{p}$ and $\Omega$ can be substituted by $\lambda_\text{p}\pm 180\degr$ and $\Omega \pm 180 \degr$ without affecting the Stokes parameters. 
Furthermore, at low eccentricities (such as the best fit values that we obtained), $\lambda_\text{p}$ and $\phi_\text{p}$ are correlated (see Eqs.\,\ref{eq:eccent_anom}, \ref{eq:meananom}, and \ref{eq:phiorb}). This effect can be observed in the joint posterior distribution of $\lambda_\text{p}$ and $\phi_\text{p}$ shown in Fig.\,\ref{fig:posteriors}.

As it follows from the formulae given in the Appendix of the paper by \citet{Simmons1984},\footnote{The paper by \citet{Simmons1984} contains an error in equation (A2): inside the last square brackets, the "+" sign is missing between the terms $2e\cos(\lambda - \lambda_\text{p})$ and $(e^2/2) \cos[2(\lambda - \lambda_\text{p})]$.} a large orbit eccentricity should lead to the appearance of a noticeable third harmonic of the orbital period in the variations of $q$ and $u$. However, in the observed polarization variability of \lsi, the contribution from this harmonic is insignificant. A good fit can only be obtained if we assume a nearly circular orbit (producing a strong second harmonic). Thus, the behavior of the variable component of linear polarization in \lsi\  is in an apparent contradiction with a high ($e \geq 0.5$) value of the eccentricity of the binary orbit.

We note that the $\chi^2$ value of the fit with constrained $\Omega$ seems too large (67 for 52 dof).
Therefore, at the next step, we relaxed the constraints on $\Omega$, choosing a very wide prior interval.
The best fit parameters together with the values of $\chi^2$ and the dof are shown in the lower part of Table \ref{tab:tbl_ecc_fit}.  
The posterior distribution is shown in Fig.\,\ref{fig:posteriors_free}. 
The comparison of the model to the data is shown in Fig.\,\ref{fig:Stokes} with the blue dotted line. 
A sketch of the possible orbit is pictured in the right panel of Fig.\,\ref{fig:orbital_elems}.  
We note that the $\chi^2$ value of the fit is much smaller than in the case of constrained $\Omega$ (50 versus 67) and the best-fit $\Omega\approx120\degr$, which differs by 90\degr\ from the value obtained in the first case.
Such a value for $\Omega$ can be interpreted in two different ways. 
This might indicate that the orbit of the compact object is nearly perpendicular to the disk plane, which is not likely (see Sect.\,\ref{sec:pol_var}). 
A more probable interpretation is that the projection of the decretion disk axis makes a 90\degr\ angle to the average intrinsic PA, and the orbit of the compact object is nearly co-planar to the disk (i.e., $\Omega\approx \theta_\text{int}+ 90\degr$).
In this geometry, the polarization vector of radiation scattered in a cloud is predominantly perpendicular to the PA of the average polarization associated with the disk.
This explains why, at the phases where the scattering angle is about 90\degr\ and maximum polarization of the scattered radiation is expected (i.e.,  at $\phi\approx0.2$ and 0.7), the minima are observed in both $q$ and $u$. 
Despite a very different geometry, the best fit eccentricity is identical to the case of $\Omega \approx \theta_\text{int}$. 
Also the best fit phase of the periastron $\phi_\text{p}$ is still very close to 0.6. 
Thus, we conclude that the obtained values for the eccentricity and the periastron phase are robust and do not depend on our assumptions on the disk and compact object orbit orientations.
We discuss the implications of the obtained orbital parameters in Sect.\,\ref{sec:orbit}.

\section{Optical photometry}
\label{sec:phot}

In addition to the optical polarimetry, we also studied the brightness variations of \lsi\  in the \textit{B}, \textit{V},  and \textit{R}  passbands by measuring the ratio of the fluxes from the binary and the nearest field star \#4. For the images taken with the UH88 telescope, the small size of the field of view places star \#4 too close to the edge of the image field, preventing us from using these images for reliable relative photometry. Thus, only the images taken with the T60 telescope have been used. An example of the light curve in the \textit{V} band, obtained from three years of observations, is shown in  Fig.\,\ref{Fig:Timeline}.

\begin{figure}
\centering
\includegraphics[width=0.80\hsize]{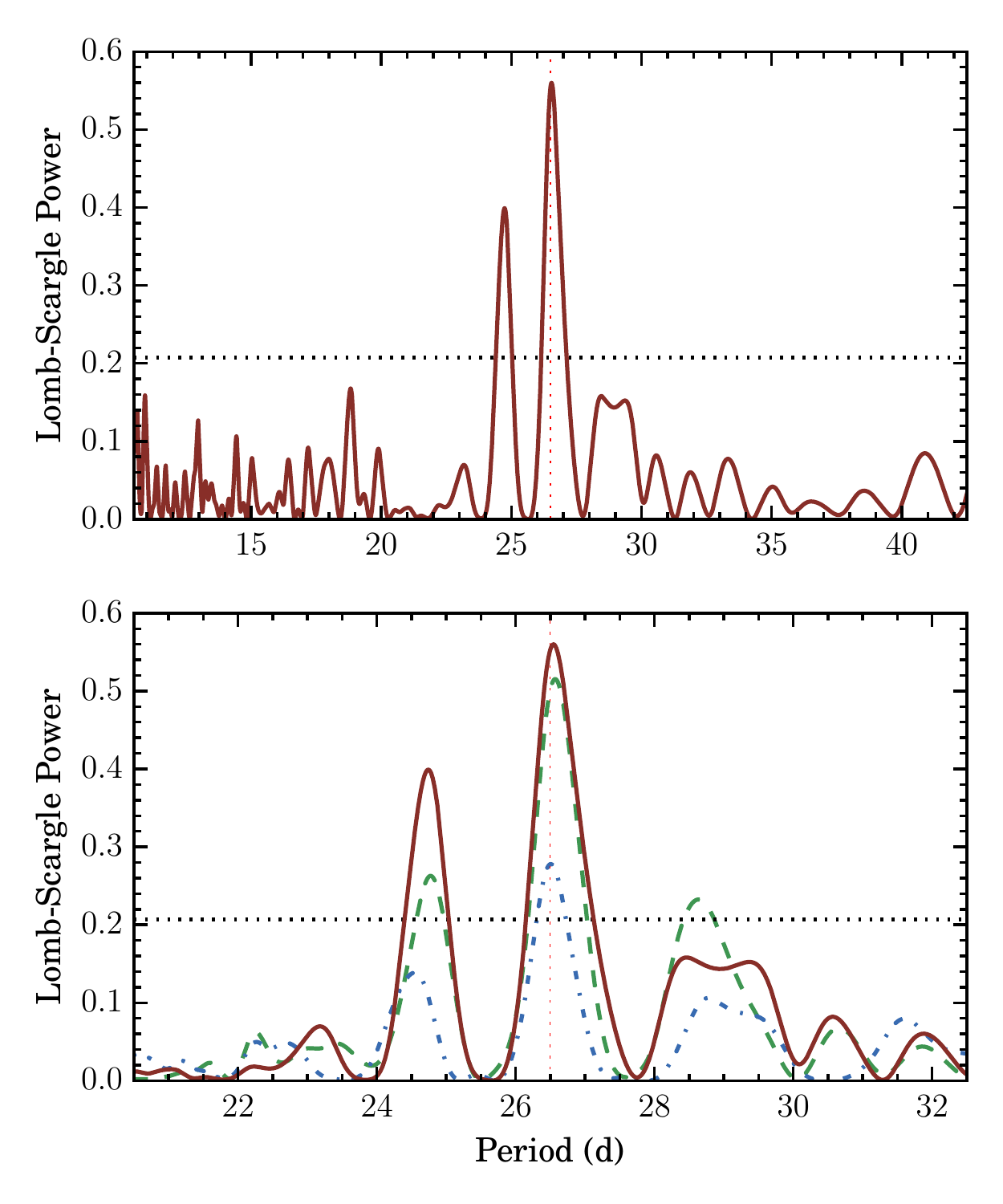}
\caption{\emph{Top panel}: Lomb-Scargle periodogram for brightness variability of \lsi\  in the \textit{R} band.
\emph{Bottom panel}: Lomb-Scargle periodogram zoom for brightness variability in the \textit{BVR} bands (dotted blue, dashed green, and solid red lines, respectively). The horizontal dotted line corresponds to the \text{FAP} = 1\%. The vertical red dotted line marks the orbital period $P_\text{orb} = 26.496$\,d.} 
\label{Fig:ScarglePhot}
\end{figure}

 As in the case of the Stokes parameters, we used a Lomb-Scargle timing analysis to find the period of the variability of the brightness in \lsi. The frequency of the highest peak on the Lomb-Scargle periodograms  (Fig.\,\ref{Fig:ScarglePhot}) is close, but not exactly equal to the orbital period  $P_\text{orb} = 26.4960 \pm 0.0028$\,d. For the \textit{BVR} bands, the determined periods are $P_B = 26.56 \pm 0.05$, $P_V = 26.58 \pm 0.04$, and $P_R = 26.60 \pm 0.03$\,d, with the false alarm probabilities $\sim$ $10^{-5}$, $10^{-10}$, and $10^{-14}$, respectively. The errors of the periods were estimated via the randomization of the data within $6\sigma$ interval and recalculation of the periodogram (number of recalculations $N = 1000$).

\begin{figure}
   \centering
    \includegraphics[width=0.80\hsize]{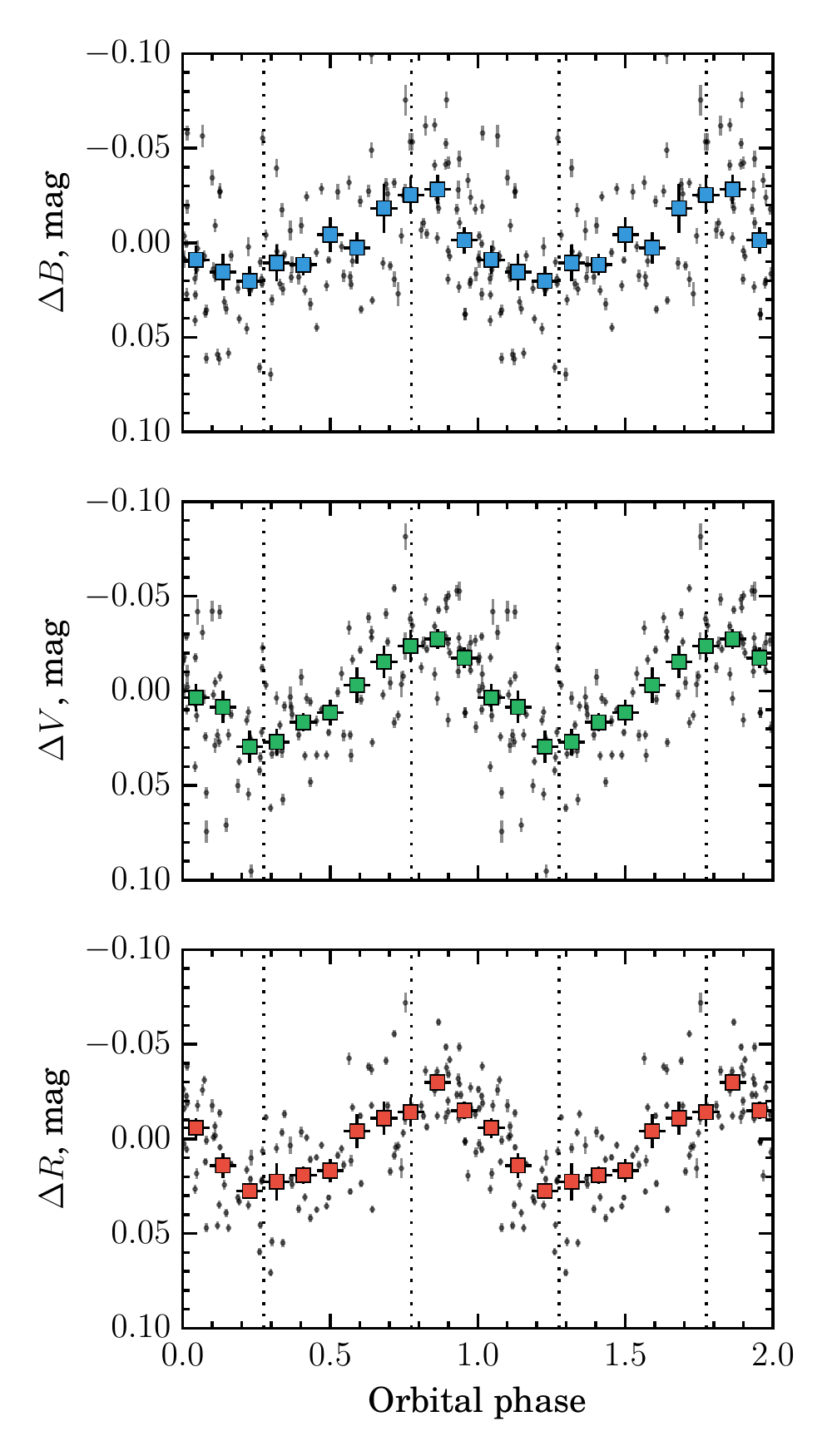}
      \caption{Variability of \lsi\  brightness in \textit{BVR} bands (\emph{top, middle, and bottom panels}, respectively) with the orbital period. The filled squares with 1$\sigma$ errors correspond to the average values of the individual observations (gray crosses) and the standard errors of the mean calculated within the phase bin of width $\Delta \phi = 0.091$.}
       \label{fig:phot_all}
\end{figure}

 The light curves of \lsi, folded with the orbital period $P_\text{orb}$\ are shown in Fig.\,\ref{fig:phot_all}. The shape of these curves (sinusoidal-like wave with an amplitude of about 0.03\,mag, minimum near the periastron and maximum in the apastron) is in a good agreement with the previous results obtained by \citet{Mendelson}, \citet{Zaitseva}, and \citet{Paredes15}. Like in the case of polarization variability, there is a noticeable scatter around the light curves in all passbands, which indicates the presence of a nonperiodic component.

\begin{figure}
    \centering
    \includegraphics[width=0.8\hsize]{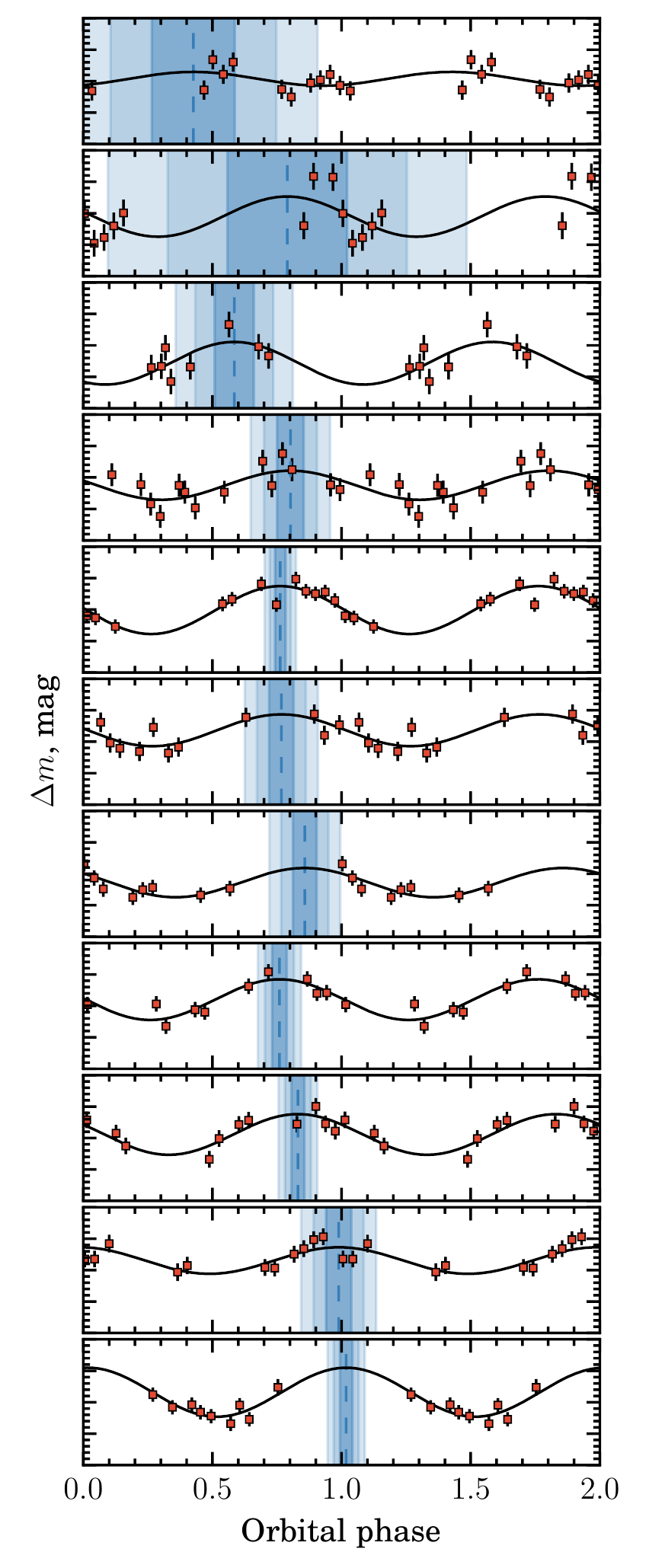}
      \caption{Variability of \lsi\ optical brightness\  in \textit{V}-band for different parts of the data (seasons S1--S11  \emph{from top to bottom}). The black solid lines correspond to the best fit of the data with function  (\ref{eq:cos}). The vertical dashed lines give the phases of the best fit maxima $\phi_0$, and 
       the corresponding $\pm1\sigma$, $\pm2\sigma$, and $\pm3\sigma$ confidence intervals are shown with varying shades of blue. The vertical scale [0.1, $-$0.1] is the same in all panels. }
       \label{fig:phase_shift}
\end{figure}

\begin{table}   
\caption{Parameters of the best fit to the optical photometry of \lsi\  in the \textit{V} band with function (\ref{eq:cos}) for eleven observing seasons. } 
 \label{table:phot}      
\centering                          
\begin{tabular}{l c r@{$~\pm~$}l c r@{$~\pm~$}l}        
\hline\hline                 
Seasons & MJD & \multicolumn{2}{c}{$A$}  & $\phi_0$ & \multicolumn{2}{c}{$m_0$}  \\    
 &   & \multicolumn{2}{c}{(mmag)}  &  & \multicolumn{2}{c}{(mmag)} \\
\hline                        

S1 & 57687--57702  & 11 & 6  & $0.42 \pm 0.16$ &  $-$4 &  6\\
S2 & 57777--57785  & 32 & 17 & $0.78 \pm 0.23$ &  5  & 40\\
S3 & 58026--58054  & 34 & 22 & $0.58 \pm 0.08$ &  28 & 15\\
S4 & 58056--58082  & 23 & 6  & $0.80 \pm 0.05$ &  12 & 5\\
S5 & 58092--58117  & 38 & 7  & $0.76 \pm 0.02$ &  1  & 5\\
S6 & 58125--58159  & 25 & 7  & $0.76 \pm 0.05$ &  $-$18 &6\\
S7 & 58337--58352  & 23 & 9  & $0.85 \pm 0.05$ &  14 & 6\\
S8 & 58407--58429  & 32 & 6  & $0.75 \pm 0.03$ &  $-$10 & 4\\
S9 & 58456--58474  & 32 & 7  & $0.83 \pm 0.03$ &  $-$6 & 4\\
S10 & 58684--58709 & 21 & 6  & $0.98 \pm 0.05$ &  $-$15 & 4\\
S11&  58715--58728 & 39 & 11 & $1.01 \pm 0.03$ &  $-$16 & 8\\

\hline                                   
\end{tabular}
\end{table}

 In order to find a possible superorbital phase shift of the maximum of brightness, which is discussed for example in \citet{Paredes15}, we divided our data into eleven subsets (see Table \ref{table:phot} and Fig.\,\ref{Fig:Timeline}). We then folded them with the orbital period $P_\text{orb}$ and fit with the function 
 \begin{equation} \label{eq:cos}
 m(\phi)=-A\cos[2\pi(\phi-\phi_0)]+ m_0, 
 \end{equation}
 where $\phi_0$ (phase of the peak), $A$ (amplitude), and $m_0$ (vertical offset) are free parameters. 
 The data and the best fit curves are shown in Fig.\,\ref{fig:phase_shift}.
 The values of the best fit parameters are given in Table \ref{table:phot}. 
 The errors on the parameters were calculated as a square root of the diagonal elements of the covariance matrix of the fit. The measurement errors are smaller than the intrinsic scatter in the photometric data. For a more accurate estimation of errors of the fit parameters, we used the standard deviation of the data instead of the measurement errors for the fitting (in that case, the reduced $\chi^2 \approx 1$). 

\begin{figure}
\includegraphics[width=0.85\hsize]{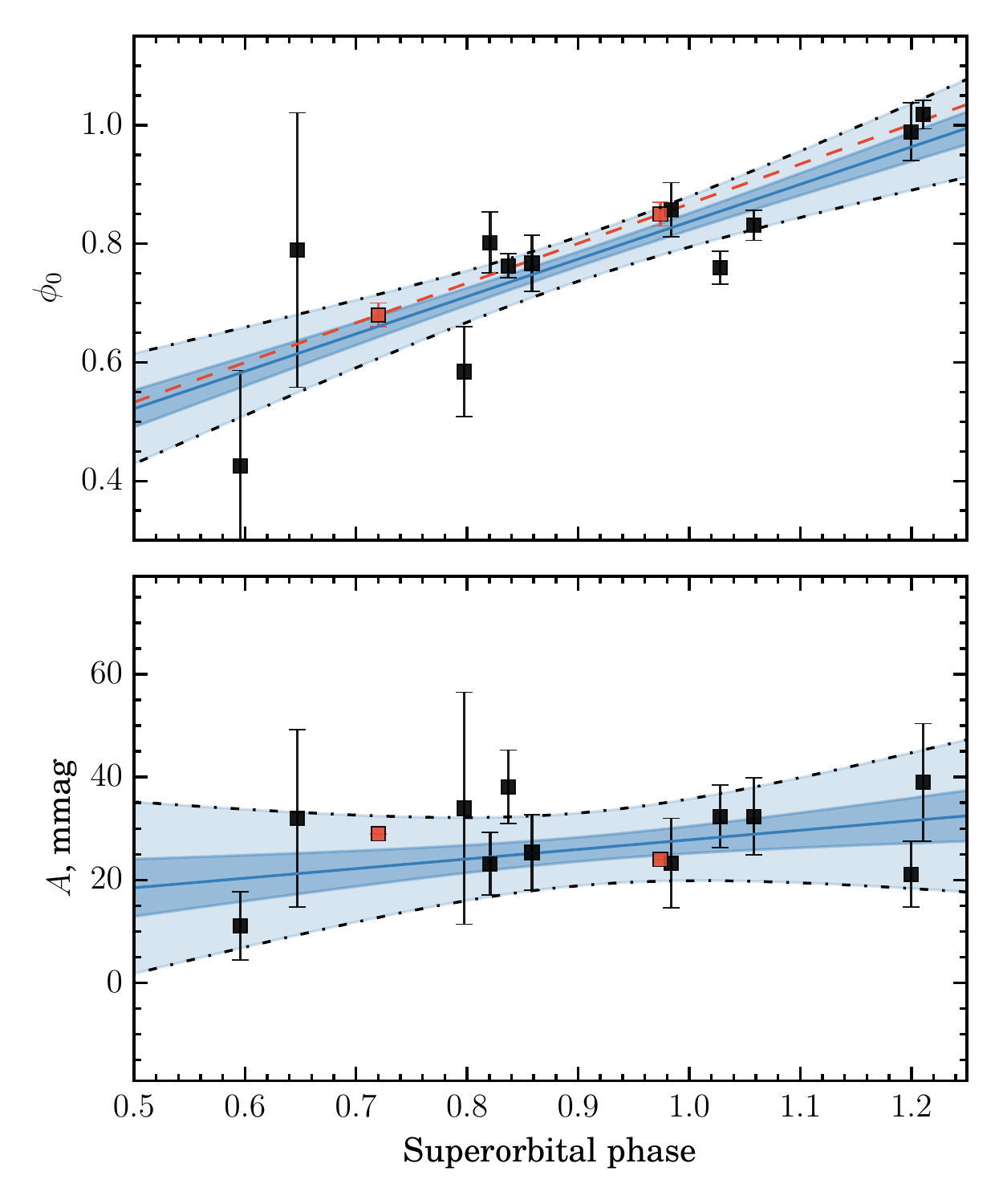}
\caption{Dependence of brightness maximum orbital phase  $\phi_0$ (\emph{top panel}) and the amplitude $A$ (\emph{bottom panel}) of the sinusoidal fits on the superorbital phase. 
The solid blue lines correspond to the linear fit of the data, while the $\pm1\sigma$ and $\pm3\sigma$ confidence intervals  are shown in dark and light blue. The red squares show the parameters from Table 3 of \citet{Paredes15}.}
\label{fig:phase}
\end{figure}

We see from Fig.\,\ref{fig:phase_shift} that there is a significant phase shift by $\Delta \phi_0 \approx 0.3$ seen between observing seasons S1 and S11. The quality of fits for the first two seasons, S1 and S2, are quite low, but the linear shift of the phase of the maximum of brightness is apparent even without them. Thus, our new photometry data are in agreement with the results obtained by \citet{Paredes15}. The relatively short time baseline (three years) does not allow us to conclude whether this shift is periodic (superorbital) or linear. We have also studied a dependence of the amplitude of brightness variability over time (see bottom panel of Fig.\,\ref{fig:phase}). As is seen, there is no apparent trend here, unlike the one seen in the phase of the maximum. We must emphasize that, due to the presence of the nonperiodic brightness fluctuations, studying possible periodic (superorbital) events is complicated. These fluctuations introduce significant noise, which may completely diminish small-amplitude gradual changes occurring over long time scales. 

It is worth pointing out that the observed light curve does not necessarily imply a high eccentricity of the binary orbit. On the contrary, if the  brightness variations result from disruption of the disk by the compact object at the periastron passage \citep[see][]{Paredes15}, one would expect light curve asymmetry, that is, a fast dip near the previously assumed periastron phase $\phi_\text{p} \simeq 0.275$ with more gradual growth toward apastron at $\phi_\text{a} \simeq 0.775$. The observed light curves in all passbands, apart from being noisy, do not show such asymmetry. The small and nearly equal amplitude in all passbands and the shape of light curve variability, can be similarly (if not better) explained by the occultation of the Be star or a bright emission area, by a dense gaseous cloud. The brightness variability can be interpreted in terms of viewing geometry, and this explanation does not require an eccentric orbit and periodic disk disruption. 
We note that the phase of the maximum brightness, $\phi_0 = 0.78 \pm 0.15$, differs significantly from the previously assumed periastron phase $\phi_\text{p} = 0.275$, it is, however, much closer to our estimate of the periastron phase of $\approx$0.6. 
We discuss this fact in Sect.\,\ref{sec:periastron}.

\section{Orbital parameters of \lsi}
\label{sec:orbit}

\subsection{Eccentricity}
\label{sec:eccenticity}

The estimates of the orbit eccentricity $e$ derived for \lsi\  by different methods, vary across a wide range from 0.3 to 0.8 \citep[see][and references therein]{Grundstrom2007}. The most reliable and direct way to reconstruct the binary orbit is to measure radial velocity (RV) variations of the stellar atmospheric spectral lines. Several efforts have been made to find orbital parameters for \lsi\  from the RV variations \citep{Hutchings1981, Casares2005, Grundstrom2007, Aragona2009}. The latest solutions obtained from the RV curves for the \ion{He}{I} and \ion{He}{II} lines give the value of eccentricity from $\simeq0.54$ \citep{Aragona2009} to $\simeq0.72$ \citep{Casares2005}.

The common feature seen in all RV curves for the \lsi\  is a high degree of scatter of individual measurements around the fitting curve, which is particularly pronounced in the orbital phases after the previously assumed periastron phase around 0.3 and the 0.8--0.9 phase range. Another feature is the apparent presence of a ``secondary bump'' near the phase 0.7--0.8   \citep{Grundstrom2007, Aragona2009}. 

We want to emphasize that such a shape of the RV curve is not exceptional, but rather typical for many Be components in X-ray and non-X-ray binary stars. The formal solution of such a curve often results in substantial eccentricity for the Be star orbit. An explanation of this  phenomenon was proposed by \citet{Harmanec1985} (see his Figs.\,1 and 3).\footnote{The alternative model of massive X-ray binaries suggested by \citet{Harmanec1985}, which assumes that X-ray components in massive systems are not neutron stars or black holes, is outdated. However, his interpretation of the RV curves of spectral lines formed in the vicinity of the visible component in Be binary systems is correct at least in some cases.}
The optical component in most of such systems is embedded in a dense disk. The spectral lines, seen in the optical and UV spectra and usually associated with this star, show complex profiles, often with emission wings. From that point of view, \lsi\  is a typical Be-type binary with spectral lines closely resembling so-called shell lines, which are identified in spectra of many interacting binary stars \citep[see binary spectra published in][]{Casares2005, Grundstrom2007, Aragona2009}. Even the lines that do not show prominent emission are not purely atmospheric, but may be formed in the different parts of the optically thick disk or gaseous shell around the star. The RVs, measured for these lines (apart from the orbital motion) are affected by the complex kinematics of the gas motion in the different parts of the disk. Thus, they may not adequately reproduce the orbital motion of the Be star, and the solution of RV curves may result in spurious (and different for the different lines) values of the eccentricity of the orbit (\citealt{Harmanec1985}; see also more recent papers by \citealt{Harmanec2015} and \citealt{Koubsky2019}).  

The complications with RV curves are well illustrated by the behavior of the shell lines in the Be binary star KX And \citep{Stefl1990}. The formal solutions of the RV curves obtained for the six different groups of lines formed in the vicinity of the primary give an orbit eccentricity value ranging from 0.22 to 0.64 \citep[see Table V in][]{Stefl1990}. In contrast, the RV variations of the spectral lines of the late-type secondary, which were detected and studied with high-resolution spectroscopy by \citet{Tarasov1998}, revealed zero eccentricity and proved circular orbit in KX And. 

We believe that the high value of eccentricity $(e \geq 0.5)$ obtained from the RV curves for the \lsi\  \citep[and perhaps for the LS 5039, see][]{Aragona2009} is spurious. Of course, there is a morphological  difference between the Be systems with normal (nondegenerate) secondary components and high-mass X/$\gamma$-ray binary systems with black hole/neutron star companions. However, the remarkable resemblance of the RV curves of the Be component in systems like KX And and many massive X/$\gamma$-ray binaries, including \lsi\  \citep[see Fig. 3 in][]{Harmanec1985}, cannot be ignored. Thus, the real orbit in \lsi\  might be only slightly eccentric and close to circular. This is also suggested by the behavior of the periodic (orbitally phase-locked) variable polarization that we observed in the system.

\subsection{Phase of the periastron}
\label{sec:periastron}

In addition to the high eccentricity, the RV measurements resulted in the value for the orbital phase of the periastron $\phi_{\rm p} = 0.22-0.23$ \citep{Casares2005}, $\phi_{\rm p} = 0.3$ \citep{Grundstrom2007}  and  $\phi_{\rm p} = 0.275$ \citep{Aragona2009}. 
However, the peak on the light curves does not occur on this orbital phase at most wavelengths. For example, the peaks in the radio, X-ray, and TeV light curves occur around the phase $\phi \approx 0.6$  \citep{Massi15,Archamault2016,Chernyakova2017}. 
The peak of the optical brightness occurs with some delay relative to the peak in the high-energy emission (see  Sect.\,\ref{sec:phot}, as well as \citealt{Zamanov2014}, and \citealt{Paredes15}). 
It is difficult to understand the nature of the phase delay between the passage of the periastron by a compact object and the maximum of radiation, especially in the case of high eccentricity. 
Our value for the orbital phase of periastron $\phi_{\rm p} \approx 0.6$ (see Table \ref{tab:tbl_ecc_fit} and Fig.\,\ref{fig:orbital_elems} for orbital parameters and a possible geometry), obtained from the modeling of the optical polarization, may explain all observational facts mentioned above as an interaction of the compact object, moving around the Be star, with the densest part of the circumstellar disk, near the periastron. 
A new set of orbital parameters should now be considered when modeling light curves in different energy bands.

\section{Summary}

Our new high-precision \textit{BVR} measurements of the linear polarization of the $\gamma$-ray binary \lsi\ revealed periodic orbital variability in all passbands. The timing analysis of the Stokes parameters yielded the first ever detection of a polarimetric period $P_{\text{Pol}} = 13.244$\,d, which is close to half of the orbital period $P_\text{orb} = 26.496$\,d. The continuous change of polarization with the orbital phase implies that the variations arise from the orbital motion of the compact star whose orbit is co-planar with the Be star's decretion disk. The mechanism producing orbital variation of the polarization  is most likely Thomson scattering of the stellar light in the high-temperature region around the compact object.
The orbital variability curve is dominated by the second harmonic, which is typical for binaries with close to circular orbit and nearly symmetric distribution of the light scattering material with respect to the orbital plane. This implies that the high eccentricity of the binary orbit in \lsi\  derived from the solution of the RV curves may not be real and that the true orbit is close to circular. 

After the determination and subtraction of the interstellar polarization component, we obtained the PA of the constant component of polarization associated with the Be decretion disk at $\theta_\text{int}\simeq11{\degr}$. Although this value differs from the previously determined $\simeq25{\degr}$ \citep{Nagae2006, Nagae2009}, we believe that the difference is most likely due to the uncertainties in the determinations of the interstellar polarization component.  

We considered two cases: when the position angle of the disk normal nearly coincides with $\theta_\text{int}$, and when it differs by 90\degr\ from it.
Using a model of a scattering cloud at an elliptical orbit nearly co-planar with the disk, we  modeled orbital variations of Stokes parameters and constrained the eccentricity at $e<0.15$ and the phase of the periastron at  $\phi_\text{p}\approx 0.6$.  
Both constraints are independent of the assumption on the disk orientation.
The longitude of the periastron was found to be $\lambda_\text{p}\approx 146\degr$ or 225\degr\ for the two cases.
The obtained value for the periastron phase differs significantly from the previously assumed phase, which is based on the RV measurements, but is very close to the peaks of the radio, X-ray, and TeV light curves.  
Our results thus open a new avenue to model the broad-band emission from the enigmatic $\gamma$-ray  binary \lsi.  

We also found photometric orbital variability of \lsi\  in \textit{B}, \textit{V}, and \textit{R} filters with amplitudes $\Delta m \approx 0.1$\,mag. The phase shift of the brightness maximum  between the data sets acquired over the period of three years can be approximated with a simple linear model.

\section*{Acknowledgements}

We acknowledge support from the Magnus Ehrnrooth foundation, the Finnish National Agency for Education (EDUFI) Fellowship (VK), the Russian Science Foundation grant 20-12-00364 (VK, SST and JP), the  Academy  of  Finland grant 333112, and the ERC Advanced Grant HotMol ERC-2011-AdG-291659 (SVB, AVB).  
We thank  the  Academy  of  Finland (projects 317552, 331951) and the German Academic Exchange  Service  (DAAD,  projects  57405000 and 57525212) for travel grants.
DM acknowledges support from the DFG grant MA 7807/2-1 and from the bwHPC project of the state of Baden-W\"urttemberg.
The Dipol-2 polarimeter was built in cooperation by the University of Turku, Finland, and the Leibniz Institut f\"{u}r Sonnenphysik, Germany, with support from the Leibniz Association grant SAW-2011-KIS-7. 
We are grateful to the Institute for Astronomy, University of Hawaii, for the allocated observing time.

\bibliographystyle{aa}
\bibliography{aanda.bib}


\appendix

\section{Polarization from Thomson scattering by an orbiting  cloud}
\label{sec:appendix}

We consider an orbital plane with the rotation axis $\unit{n}$ inclined by angle $i$ to the line of sight. 
Let us choose the polarization basis $(\unit{e}_1,\unit{e}_2)$ such that vector $\unit{e}_1$ is along the projection of the vector $\unit{n}$ on the plane of the sky, $\unit{e}_2$ lies at the interception of the plane of the sky, and the orbital plane is perpendicular to $\unit{e}_1$. 
This basis can be supplemented by $\unit{e}_3,$ which coincides with the direction to the observer $\unit{o}$  to form a right-handed coordinate system. 
In this system,  $\unit{n}=(\sin i, 0, \cos i)$.
The center of the coordinates is on the source of light (i.e., the Be star).

\begin{figure}
\centering
\includegraphics[width=\hsize]{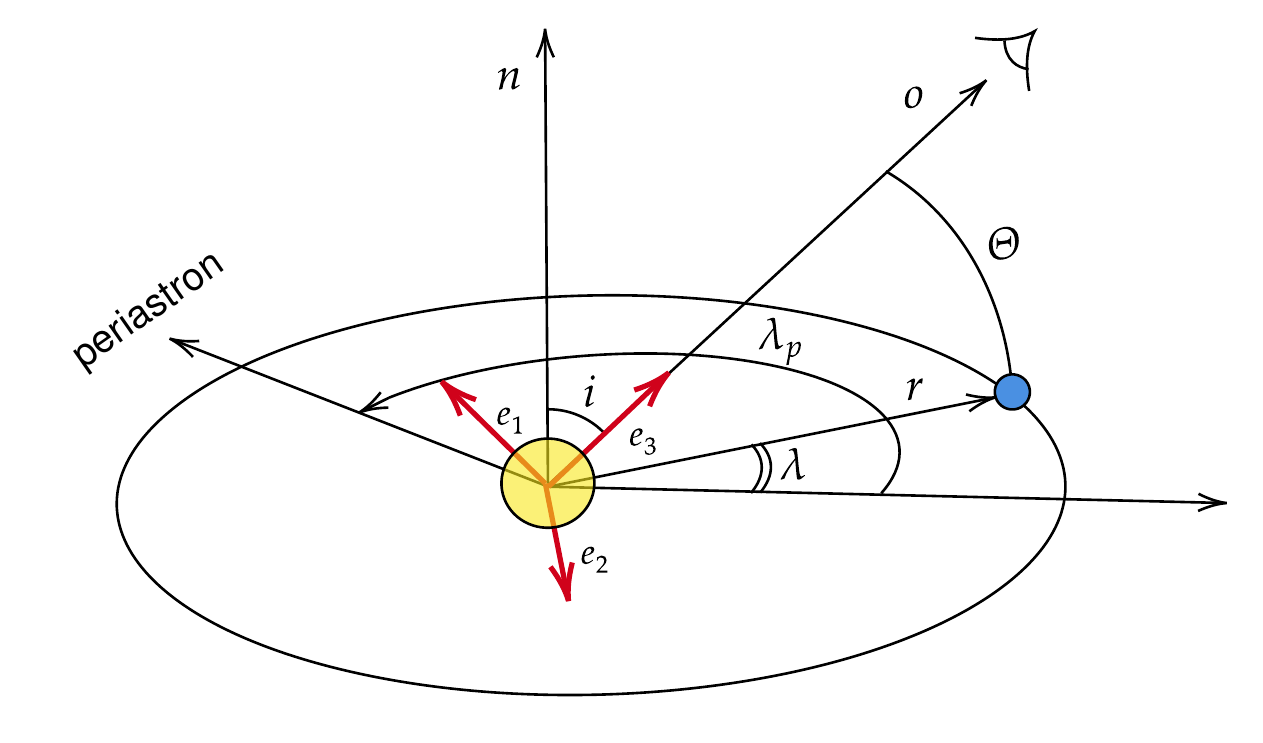}
\caption{\label{fig:orbit} 
Geometry of orbit. }
\end{figure}

Let us introduce longitude $\lambda$ in the orbital plane measured from the southern part of the line formed by the intersection of the orbital plane with the plane containing ($\unit{n}$, $\unit{o}$), as in Simmons and Boyle 1984 (see our Fig.\,\ref{fig:orbit} and their Fig.\,1).
The orbital distance to the scattering cloud varies with longitude as  
\begin{equation} 
r(\lambda)= \frac{a(1-e^2)}{1+ e \cos (\lambda-\lambda_{\rm p})}, 
 \end{equation} 
where $a$ is the orbit semi-major axis and $\lambda_{\rm p}$ is the longitude of the periastron.
The unit vector towards the cloud is
\begin{equation} 
\unit{r}= (-\cos i \cos\lambda, -\sin\lambda, \sin i \cos \lambda)  
 ,\end{equation} 
and the scattering angle $\Theta$ is given by 
\begin{equation} 
\mu= \cos\Theta = \unit{r} \cdot   \unit{o}  =  \sin i \cos \lambda . 
 \end{equation} 

The total flux observed at Earth from the system $F_{\rm tot}$ consists of the direct radiation from the Be star $F_* = L_* /(4 \pi D^2 )$ and the scattered flux $F_{\rm sc}=  L_{\rm sc}(\mu) /(4 \pi D^2 )$, where $L_*$ and $L_{\rm sc}$ are corresponding luminosities and  $D$ is the distance.
Assuming that Thomson scattering (in an optically thin regime) is responsible for  scattered flux, the angular distribution of scattered luminosity can be represented as 
\begin{equation} 
L_{\rm sc}(\mu)  = L_* \ f_{\rm sc}\  l(\mu), 
 \end{equation} 
where 
\begin{equation} 
l(\mu)= \frac{3}{8} (1+\mu^2) 
 \end{equation}  
is the Thomson scattering indicatrix, 
\begin{equation} 
f_{\rm sc} = \frac{N_{\rm e}\sigma_{\rm T}}{4\pi r^2(\lambda)} 
\end{equation} 
is the fraction of scattered photons that is related to the total number of free electrons $N_{\rm e}$ in a cloud and the distance between the Be star and the scattering cloud, and $\sigma_{\rm T}$ is the Thomson cross-section. 
We can scale $f_{\rm sc}$ to some typical value $f_0$ as
\begin{equation} 
f_{\rm sc}  = f_0\ \left[\frac{a(1-e^2)}{r(\lambda)}\right]^2 = f_0 \ [1+ e \cos (\lambda-\lambda_{\rm p})]^2 . 
 \end{equation} 
 
The polarization degree of scattered light is 
\begin{equation} 
P_{\rm sc}= \frac{1-\mu^2}{1+\mu^2} = \frac{\sin^2\Theta}{1+\cos^2\Theta}, 
 \end{equation} 
but the observed one is diluted by the star.
The polarized flux for the assumed Thomson scattering indicatrix is 
\begin{equation} 
F_{\rm sc} P_{\rm sc} = F_*\  f_{\rm sc}\ l(\mu)\   \frac{1-\mu^2}{1+\mu^2}  
= F_*\  f_{\rm sc}\   \frac{3}{8}\ (1-\mu^2) .
 \end{equation} 
The polarization degree of the total flux is $P= F_{\rm sc} P_{\rm sc} / F_{\rm tot}$, but if we assume that scattered radiation contributed very little to the total flux $F_{\rm sc} \ll F_*$, we get 
\begin{equation} \label{eq:pol_tot}
P=  \frac{3}{8}\ f_{\rm sc} \ (1-\mu^2) = 
\frac{3}{8}\ f_0 \ \sin^2\Theta \ 
[1+ e \cos (\lambda-\lambda_\text{p})]^2 .
 \end{equation} 

The polarization (pseudo-)vector $\unit{p}$, defined by the direction of dominant oscillations of electromagnetic waves, is perpendicular to the scattering plane, so
\begin{equation} 
\unit{p} =  \frac{\unit{o} \times  \unit{r}}{|\unit{o} \times  \unit{r}|} 
= \frac{1}{\sin\Theta}\left(\sin\lambda,  -\cos i \cos\lambda, 0\right) ,
 \end{equation} 
 and $\sin\Theta= \sqrt{1-\mu^2}= \sqrt{1-\sin^2 i \cos^2\lambda}$. 
 The position angle $\rchi$ of $\unit{p}$ is given by the following formulae:
\begin{eqnarray} 
\cos\rchi &=& \unit{e}_1 \cdot  \unit{p} = \frac{\sin\lambda}{\sin\Theta}, \\ \sin\rchi &=&  \unit{e}_2 \cdot  \unit{p} =  - \frac{\cos i \cos\lambda}{\sin\Theta}.
\end{eqnarray}
Because the normalized Stokes parameters are defined as $q=P \cos(2\rchi)$ and $u=P\sin(2\rchi)$, we need 
\begin{eqnarray} 
\cos(2\rchi) &=& 
\frac{\sin^2 i - (1+\cos^2 i) \cos(2\lambda)}{2\sin^2\Theta}, \\
\sin(2\rchi) &=& - \frac{\cos i  \sin(2\lambda)}{\sin^2\Theta} . 
\end{eqnarray}
Combining that with expression \eqref{eq:pol_tot} for polarization degree $P$, we get  
\begin{equation}\label{eq:model_Stokes}
\begin{aligned}
q &=  \frac{3f_0}{16}  \left[\sin^2 i - \left(1\!+\!\cos^2\! i\right) \cos2\lambda\right]\left[1\! +\! e \cos \left(\lambda\!-\!\lambda_\text{p}\right)\right]^2, \\
u &= \frac{3f_0}{8} \left[ -  \cos i\  \sin2\lambda\right] \left[1+ e \cos \left(\lambda-\lambda_\text{p}\right)\right]^2  .
\end{aligned}
\end{equation}
The observed Stokes parameters also depend on  the position angle $\Omega$ of the projection of the orbital axis, which are obtained by rotating vector $(q,u)$ by angle $2\Omega$: 
\begin{eqnarray} \label{eq:q_rotate}
q_\text{obs} &=&  q\cos(2\Omega)-u\sin(2\Omega),\\
 \label{eq:u_rotate}
u_\text{obs} &=&  q\sin(2\Omega)+u\cos(2\Omega).
\end{eqnarray}

The obtained Stokes parameters are functions of the longitude $\lambda$ and need to be computed as functions of the orbital phase. 
From the true anomaly of the orbit $\lambda- \lambda_\text{p}$, we find the eccentric anomaly 
\begin{equation} \label{eq:eccent_anom}
\tan\left(\frac{E}{2}\right)=\sqrt{\frac{1-e}{1+e}} \tan\left(\frac{\lambda- \lambda_\text{p}}{2}\right)
\end{equation} 
and the mean anomaly 
\begin{equation} \label{eq:meananom}
M=E-e\sin E ,
\end{equation} 
which can be converted to the orbital phase measured in the interval [0,1] using an additional free parameter, the phase of the periastron  $\phi_\text{p}$:    
\begin{equation} \label{eq:phiorb}
\phi_\text{orb} = M/(2\pi) + \phi_\text{p}.
\end{equation} 
Thus, the model free parameters are the inclination $i$, eccentricity $e$, latitude of the periastron $\lambda_{\rm p}$, the PA of the projection of the orbit axis on the sky $\Omega$, the phase of the periastron  $\phi_\text{p}$, and the scattering fraction $f_0$. We note that $\Omega$ defined this way differs by $\pi/2$ from the PA of the ascending node usually used to describe binary star orbits. 
In order to fit the data, there is a need for additional constant Stokes parameters $q_0$ and $u_0$, which describe permanent polarization and are not related to the orbital motion of the scattering cloud.

\end{document}